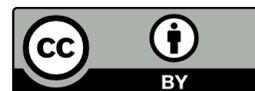

# Quantum simulator to emulate lower dimensional molecular structure

E. Sierda[1], X. Huang[1], D.I. Badrtdinov[1], B. Kiraly[1], E.J. Knol[1], G. C. Groenenboom[1], M.I. Katsnelson[1], M. Rösner[1], D. Wegner[1]*, A.A. Khajetoorians[1]*

**Affiliations:**

[1]Institute for Molecules and Materials, Radboud University, Nijmegen, The Netherlands
*corresponding authors: a.khajetoorians@science.ru.nl, d.wegner@science.ru.nl

Bottom-up quantum simulators have been developed to quantify the role of various interactions, dimensionality, and structure in creating electronic states of matter. Here, we demonstrated a solid-state quantum simulator emulating molecular orbitals, based solely on positioning individual cesium atoms on an indium antimonide surface. Using scanning tunneling microscopy and spectroscopy, combined with *ab initio* calculations, we showed that artificial atoms could be made from localized states created from patterned cesium rings. These artificial atoms served as building blocks to realize artificial molecular structures with different orbital symmetries. These corresponding molecular orbitals allowed us to simulate 2D structures reminiscent of well-known organic molecules. This platform could further be used to monitor the interplay between atomic structures and the resulting molecular orbital landscape with sub-molecular precision.

**One-Sentence Summary:** Molecular orbitals of organic molecules were simulated solely based on the use of atomically positioned cesium ions on the surface of a semiconductor.

**Introduction**

Fascinating new physical and chemical phenomena can arise when electrons in solids are confined to lower dimensions (*1, 2*). A spectacular example is graphene, in which the motion of electrons becomes relativistic, which has profound consequences on its material properties including chiral (Klein) tunneling, Zitterbewegung, relativistic collapse around charge centers, or emerging pseudomagnetic fields from inhomogeneous deformations (*3-6*). Due to confinement, mutual interactions between electrons can furthermore become significant in lower dimensional systems, often leading to new and unexpected fragile quantum states of matter, such as Mott insulators, superconductivity, or spin liquid behavior (*7-11*). Understanding quantitatively how these exotic quantum states of matter emerge requires understanding the detailed interplay between dimensionality, multi-orbital electronic structure, and interactions.

In order to provide insight into interacting lower dimensional systems and to take advantage of the resultant fragile many-body states of matter, new types of tunable quantum simulators have been developed toward the bottom-up design of many-body Hamiltonians (*12-20*). One approach toward this end is the precise atom-by-atom creation of potential landscapes on surfaces, based on scanning tunneling microscopy/spectroscopy (STM/STS) (*21, 22*). Using this approach, various artificial two-dimensional (2D) lattices have been created to emulate the single-particle band structure of Dirac and topological materials (*23-25*). While the versatility of this approach is rooted in the countless geometries that can be realized, it often relies on using supporting bulk metallic substrates. The latter is a clear drawback in designing atomic and molecular orbitals with tailored symmetry, as well as tuning mutual electron interactions, due to the unwanted influence of the substrate's bulk electronic bands. Therefore, to harness the versatility of this approach, it is vital to develop platforms that use weakly conducting or insulating surfaces (*26, 27*).

Here, we demonstrated a versatile quantum simulator capable of realizing complex artificial molecular orbitals based solely on patterning individual cesium (Cs) atoms on the surface of semiconducting indium antimonide (InSb). Using STM/STS, we first constructed building blocks from patterned ring structures of Cs atoms, which exhibited a zero-dimensional (0D) like state, attributable to an artificial atom. This sharply observable bound state in the band gap of the semiconducting substrate, was hence strongly decoupled from the bulk and exhibited spatially extended charge density outside the artificial atom. We subsequently showed that two artificial atoms exhibited a long-range distance-dependent coupling leading to the formation of bonding and antibonding states. Using this as a starting point, we demonstrated the multi-orbital nature of the simulator, based on creating artificial atomic $s$, $p_x$ and $p_y$ orbitals. We used various orbital symmetries and constructed 2D structures with defined geometries that are reminiscent of analogous planar three-dimensional (3D) molecular structures. The resultant electronic structure exhibited 2D artificial molecular orbitals, namely molecular σ and π orbitals, with various in-plane symmetries due to orbital hybridization. Using *ab initio* calculations of the analogous molecules, we compared the resultant and expected molecular orbitals. As the tailored structures are not restricted by the geometrical relaxation of their molecular analogues, this platform provides an ultimate tool to explore e.g. fragile and degenerate states and the interplay of these states with geometrical structure. We demonstrated this by exploring 2D structures that were reminiscent of various conformations of butadiene molecules.

**Artificial atoms**

InSb is a semiconductor with a band gap of 235 meV (*28*). Upon adsorption on its (110) surface, Cs atoms resided at hollow sites of the Sb-terminated surface (see Supplementary Materials, Section S1, Figs. S1,S2) (*29*). Concomitantly, charge donation from Cs atoms led to band bending and a dilute two-dimensional electron gas (2DEG), whose dimensionality has also been illustrated via the Quantum Hall effect (*30-33*). This 2DEG could be seen by a step-like onset in STS observed in the bulk band gap of the semiconductor (Fig. S3), with a carrier density of $N_{2DEG} = 2\times10^{12}$ cm$^{-2}$ where each Cs ion donated approximately 0.4 electrons (see Supplementary Materials, Section S2, Figs. S4-6), depending on the Cs density and doping of the semiconductor (*31, 32*).

We exploited the ionic nature of individual Cs atoms and patterned them into ring structures (Fig. 1A,B) in order to create a potential energy landscape that mimics the $1/r$ potential of an atom in 2D. The local concentration of Cs ions created an attractive potential for the electrons of the 2DEG, leading to a bound state within the bulk band gap that could be visualized using STS (Fig. 1C,D). This bound state manifested as a sharp peak with full width at half maximum of FWHM $\simeq$ 7 mV. Using spatially dependent d$I$/d$V$ imaging (see Supplementary Materials, Materials and Methods section) at the peak energy (Fig. 1C), which we refer to as an orbital map, we found that this bound state was not only located within the ring structure but also symmetrically delocalized up to $r = 9.0 \pm 0.4$ nm outside of the structure (see Fig. S7). The spatial delocalization of this wave function was due to the interplay of the local band bending induced by the Cs-ring structure and the weak screening of the 2DEG. As a result, we could identify this bound state as an *s*-like atomic state stemming from the circular potential, which is observable at a sufficiently large distance from the structure. We, therefore, defined this structure as an artificial atom. For the ring structures used in this manuscript, the structural deviations of the Cs atoms from circular symmetry were small compared to the length scale of the associated wave function. Therefore, such deviations led to negligible changes (e.g. anisotropy) in the electronic structure. Previously, it has also been shown that the 2DEG itself is approximately isotropic (*33*). Therefore, the resultant potential was well approximated by a circularly symmetric $1/r$ potential in the *x-y* plane at sufficient separation from the Cs-ring structure. We verified this both theoretically as well as by comparing different artificial Cs structures (see Supplementary Materials, Sections S3, Figs. S7, S8 and S4, Fig. S9, respectively). The calculations showed that the artificial atom also contained orbitals with $p_x$ and $p_y$ symmetries, which were not observed experimentally as they presumably resided above $E_F$, within the conduction band. We note that, as each Cs atom donated only a fraction of its outer 6*s* electron (*31, 32*), assigning an integer number of electrons bound to the artificial atom was not possible. As a result, the mesoscopic description of an artificial atom most likely involves many-body effects. However, for distances much larger than the Cs-Cs distance within the artificial atom, we reduced the many-body problem into an effective single-particle Hamiltonian and the aforementioned $1/r$ potential, as shown in the next section.

**Coupling artificial dimers**

In order to tailor Hamiltonians that resemble molecular orbitals, it was first necessary to demonstrate that these artificial atoms can couple. We correspondingly created dimers of two artificial atoms separated by $d \approx 11$ nm (Fig. 1E), where *d* is defined as the center-to-center distance. In Fig. 1F, we illustrate a false-color plot of spatially dependent STS taken at points along the line indicated in Fig. 1E. The spectra in Fig. 1F revealed two sharp low-energy states, separated by $\Delta E_{12} = e\Delta V_{12} = 46.0 \pm 0.6$ meV. While the lowest energy state was most strongly localized in between the two artificial atomic sites, the higher energy state exhibited a pronounced node with negligible intensity at the same location. Orbital maps (inset in Fig. 1F) confirmed that these energy levels could be identified as bonding ($\sigma$) and antibonding ($\sigma^*$) orbitals resulting from the coupling between the two artificial atomic $s$ orbitals. This behavior persisted for different spacings $d$, where the value of $\Delta E_{12}$ monotonically decreased for increasing $d$ and was non-zero up to $d \approx 25$ nm (Fig. 1G). This distance dependence could be fitted and well described by a simple tight binding model considering nearest neighbor interactions and isotropic atomic orbitals. In comparison to previous observations on other surfaces (*23-27, 34*), our data showed the sharpest peak widths with FWHM $\simeq 8$ mV, much smaller than $\Delta E_{12}$. For a detailed discussion see Supplementary Materials (Section S5, Fig. S10).

**Higher orbital symmetries in artificial atomic chains**

Before simulating molecular orbitals, it was necessary to confirm that the coupling of artificial atomic states also leads to orbitals with distinct symmetry, e.g., with $\sigma$ and $\pi$ characters. Therefore, we fabricated a linear molecular chain composed of six artificial atoms, where each artificial atomic site is separated by $d \approx 12.8$ nm (Fig. 2A). Using d$I$/d$V$ spectra taken along a line near the structure as input (see Fig. S11), more states separated by smaller energies were identified. To uncover the nature of these states, we acquired orbital maps (Fig. S12). In the following, we provided a summary, and reference Supplementary Materials (Section S6) for a more detailed discussion.

The lowest six states resembled a series of σ molecular orbitals derived from a superposition of the artificial atomic *s* orbitals. Following a textbook example of the linear combination of atomic orbitals (LCAO), an increasing number of nodes was found with increasing energy. We note that a simple tight binding model using the same parameters as for the dimers and considering only nearest-neighbor hopping was sufficient to reproduce the energies of all σ orbitals. At higher energies ($V_S \geq -51$ mV), we found three additional states (Fig. 2D-E and Fig. S12I) that we identified as a series of molecular orbitals derived from LCAO-like superpositions of artificial atomic $p_y$ orbitals, leading to artificial molecular π orbitals with a node plane along the chain axis and increasing local density of states (LDOS) above and below the chain. Again, the number of nodes along the *y*-axis increased with increasing energy. While there should still be three more $p_y$-derived molecular orbitals at higher sample biases, we did not observe them, presumably since those were located at energies where hybridization with the conduction band may have weakened and broadened their LDOS. However, we found another σ orbital at $V_s = -18$ mV (Fig. 2F) which perfectly resembled the LCAO of artificial atomic $p_x$ orbitals with bonding character, leading to increased LDOS along the chain axis between each of the artificial atomic sites.

As the form of the potential into the bulk of the InSb can be approximated as a triangular well potential (*30*), artificial atomic orbitals of $p_z$-like character cannot be emulated, limiting the dimensionality that can be explored. Accordingly, we did not discover any orbital of $p_z$-like character in any of our artificial molecular structures. Note that in 2D, $p_z$ orbitals project onto *s* orbitals, which is why many electronic properties of 2D materials originating from $p_z$-derived bands should also be observable in a pure 2D system. Moreover, evidence for π-like orbitals has previously been reported in an artificial structure (*35*), albeit orbital maps, as done here, are necessary to distinguish different orbital symmetries.

**Artificial hybrid molecular orbitals**

As a final step toward emulating molecular orbitals, it was necessary to illustrate that artificial atomic orbitals can hybridize. We therefore subsequently created a six-fold symmetric hexagon of artificial atoms, each composed of six Cs atoms, reminiscent of a benzene ring (Fig. 3A; see Fig. S14 for structural details). We chose to use a 6-Cs structure, to permit faster assembly of artificial structures containing many artificial atoms, noting it also exhibited an *s*-like state and qualitatively did not differ from the 8-Cs structure (see Supplementary Materials, Sections S4 and S11, Figs. S9, S31 and S32). In order to identify resonance energies, we acquired STS spectra at various locations (Fig. S15), and subsequently imaged the resultant orbital maps (*36*). An excerpt of maps is presented in Fig. 3 and compared with density functional theory (DFT) calculations of the actual benzene molecule. For a detailed identification and discussion of all states, we refer to Supplementary Materials (Section S7, Figs. S13-S16). In short, for energies below $E_F$, we could connect all but one orbital maps with the first eight calculated benzene valence molecular orbitals (VMOs), by comparing the experimental d$I$/d$V$ intensities with the calculated charge densities at scaled distances. In cases of degenerate VMOs we compared their superposition with the experimental maps. The overall qualitative agreement between the orbital maps and the calculated molecular orbitals provided evidence that the artificial atomic *s*, $p_x$, and $p_y$ orbitals have undergone $sp^2$ hybridization to form artificial molecular orbitals.

The versatility of this platform allowed us to create lower dimensional structures with fragile low energy states and interrogate the role of geometry on these states. While the simulated structures are not one-to-one identical to their molecular analogues (e.g. see discussion regarding $p_z$ orbitals above), studying this interplay is not easily possible in real molecular systems due to other restricting effects, such as relaxation. As an example of this versatility, we built 2D structures which were inspired by real 3D molecules that have nearly degenerate ground states and are hence unstable, and we constructed and compared 2D analogs of conformational isomers. We subsequently present examples inspired by various structural realizations of butadiene. For simplicity, we refer to these structures as artificial butadiene.

We started by coupling four artificial atoms separated by $d \approx 11$ nm that form a nearly square arrangement (Fig. 4A; see Fig. S19 for details of the structure), resembling the core of cyclobutadiene (Fig. 4B). From simple molecular orbital theory and Hückel's rule, cyclobutadiene in a fourfold symmetric geometry would be anti-aromatic and unstable, and it undergoes a Jahn-Teller distortion (*37, 38*). We applied the same approach as used for artificial benzene above to identify and map the artificial molecular orbitals, which we compared with DFT-calculated orbitals of actual cyclobutadiene in its fourfold-symmetric form (Fig. 4C-H). For a detailed identification and discussion of all states, we refer to Supplementary Materials (Section S8, Figs. S18-S22). In short, we were able to assign the first five orbital maps to the first six calculated VMOs of cyclobutadiene. Note that the maps at -67 mV and -49 mV exhibited some deviations, particularly a blurred distribution with reduced intensity. We speculate that this effect can be caused by the orbitals coupling to the 2DEG, which would delocalize the LDOS. Furthermore, small asymmetries were visible, presumably due to the fact that the structure deviated from a perfect fourfold symmetry (see Fig. S19), but with no noticeable anisotropy on the orbital levels (Fig. S20). Our simulator allows us to explore the interplay of the resultant electronic structure of analogous 2D structures, with or without such distortions.

In a similar fashion as for the cyclobutadiene structure, we also built artificial structures inspired by the *cis*- and *trans*-conformers of 1,3-butadiene, as schematically shown in Fig. 5A (for details see Supplementary Materials, Section S9-10, Figs. S23-30). We first built the *cis*-conformer (Fig. 5C) and identified all orbital maps from STS and comparison with calculations. Then, we repositioned the same group of Cs atoms of the left-hand artificial atom to build the *trans*-conformer (Fig. 5D), and we repeated all spectroscopic measurements and calculations. This way, as we built both structures using the same Cs atoms on the same surface with the same local environment, we ensured that the acquired energies are directly comparable. The schematic energy-level diagram of all nine molecular orbitals found below $E_F$ (Fig. 5B) shows that the levels of the *cis*-conformer overall tended to be a bit lower, leading a reduced sum of the occupied state energies $E_{total}^{cis}$ = -682 meV compared to $E_{total}^{trans}$ = -665 meV. In the real 1,3-butadiene molecule, the *cis*-conformer has a higher energy due to steric hindrance, leading to structural bending out of the 2D plane (*39*), emphasizing both limitations and differences between our lower dimensional platform and analogue 3D molecules. However, as an advantage, this platform would enable to change the angle $\varphi$ quasi-continuously, allowing us to experimentally map the potential landscape against this "reaction coordinate," hence providing a quantitative Walsh diagram for our artificial structure (*40*). Furthermore, equal *vs.* alternating bond lengths can be compared, in order to understand the impact of bond order in this artificial platform.

To showcase the versatility and expandability of our quantum simulator, we also built larger artificial structures representing a transition toward extended lattices. As an example, we manipulated 132 Cs atoms into a network of 22 artificial atoms arranged in a structure that is reminiscent of triangulene (see Supplementary Materials, Section S12, Figs. 32-S36), a molecule that is difficult to synthesize and has most recently drawn attention owing to its unusual electron and spin configurations (*41-46*).

There are a number of interesting observations that raise questions about the respective physical origin. For example, there is an uncanny resemblance between VMOs of organic molecules and the 2D orbital maps of the artificial structures built here. In the same vein, there is an open question as to which artificial site can be related to an actual element in terms of electron occupation (e.g. C or H) and whether artificial heterostructures (i.e. resembling different elements) can be realized. Related to this, the link between gating these artificial structures and emulating charge transfer may show exciting prospects. Finally, there are distinct differences related to the dimensionality of the platform, namely that it is below 3D, and comparisons to 3D molecular structure. These points are discussed further in Supplementary Materials (Section S13 and S14, Figs. S37, S38)

**Conclusion**

In conclusion, we have developed a versatile solid-state quantum simulator, based on using Cs ions embedded in a 2DEG on the surface of semiconducting InSb(110). We showed that artificial atoms can be created with an electronic signature given by a sharp bound state in the semiconductor band gap. The coupling of the generated electronic states with the bulk is strongly suppressed, leading to charge densities that are spatially delocalized over distances much larger than an expected atomic wave function. This leads to long-range coupling and the ability to create artificial atomic and molecular orbitals with different symmetries. Based on this, we created a variety of lower dimensional structures, from dimers to chains and eventually structures that resemble common organic molecules and characterized their low-energy electronic structure. We found an uncanny resemblance between the expected in-plane VMOs, as calculated with DFT of the free-standing molecule, and the measured orbital energic order and structure of the artificial molecules. Moreover, we illustrated that well-known conformational isomers can be artificially simulated, and the geometry perturbed, where the changes in electronic structure as a function of structural change, can be monitored. The versatility of the platform is based on being able to create fragile structures with degeneracy, that are often dominated by structural relaxations (e.g. Jahn-Teller distortion), and probe the interplay between geometry and the resultant electronic structure. This approach is distinctly different from previous studies of artificial lattices where utilized states are strongly coupled to the bulk (*21-25, 34, 47*), for example using scattered quasiparticles from patterned defects on metallic surfaces. At large distance, the attractive potential can be approximated as an atomic-like $1/r$ potential, but locally, the description involving the Cs ions and local 2D carriers most likely involved many-body interactions (*48*). In this way, due to the interplay of poor screening, lower dimensionality, and low carrier density, this platform may also be suitable for exploring tailored many-body states in designed structures, based on strong electron-electron interactions.

**Acknowledgements**

**Funding:** The experimental part of this project was supported by the European Research Council (ERC) under the European Union's Horizon 2020 research and innovation programme (grant no. 818399). We also acknowledge support from the NWO-VIDI project 'Manipulating the interplay between superconductivity and chiral magnetism at the single-atom level' with project no. 680-47-534. B. K. acknowledges the NWO-VENI project 'Controlling magnetism of single atoms on black phosphorus' with project no. 016.Veni.192.168. The work of M.I.K. was supported by the European Research Council (ERC) under the European Union's Horizon 2020 research and innovation programme, grant agreement 854843-FASTCORR. This publication is part of the project TOPCORE (with project number OCENW.GROOT.2019.048) of the research programme Open Competition ENW Groot which is (partly) financed by the Dutch Research Council (NWO).

**Author contributions:** E.S., X.H., B.K., and E.J.K. performed the experiments. D.I.B., M.I.K., and M.R. performed the DFT and model calculations, with input from G.C.G. including supplemental quantum chemical calculations. E.S., X.H., D.I.B., G.C.G., M.I.K., M.R., D.W., and A.A.K. performed and discussed the data analysis. G.C.G., D.W., and A.A.K. designed the experiments. All authors contributed to the writing of the manuscript and scientific discussion.

**Competing interests:** Non declared.

**Data and materials availability:** All data needed to evaluate the conclusions in the paper are present in the paper or the Supplementary Materials. Data for all figures presented in this study are available at Radboud Data Repository (*59*). The following software was used: Gwyddion 2.60 and Matlab R2022a (data processing), VESTA 3.5.7 (visualization of DFT calculations), Blender 3.3.1 and Adobe Illustrator 27.1 (figure preparation and 3D rendering).

**Supplementary Materials**

Materials and Methods

Supplementary Text

Figs. S1 to S38

Table S1

References (49-58)

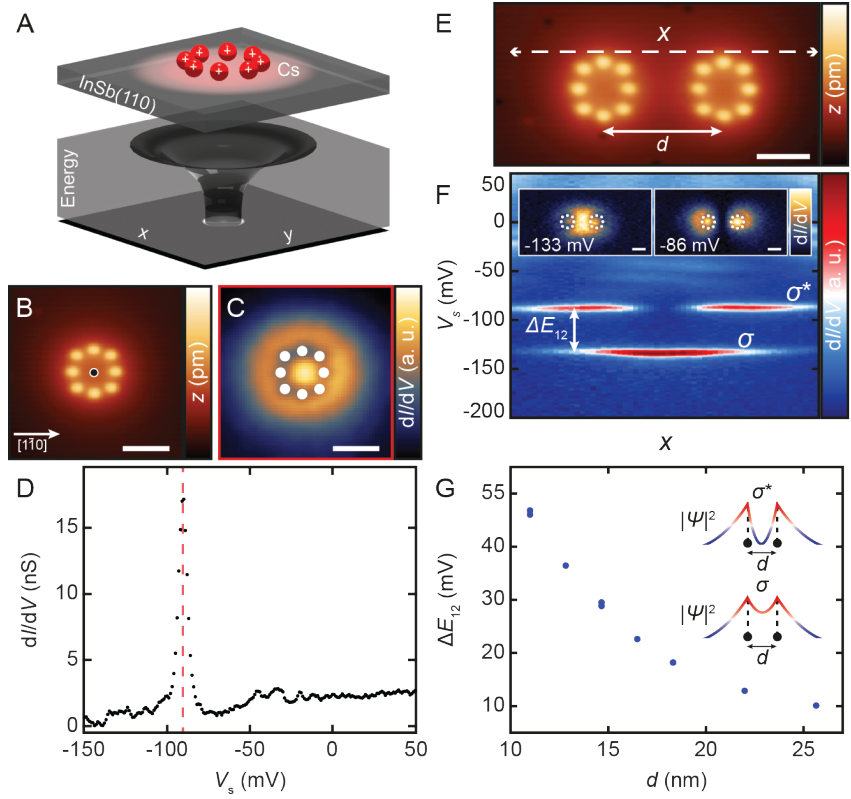

**Fig. 1. Formation of an artificial atom from Cs atoms on InSb(110) and bonding/antibonding orbitals of coupled artificial atoms.** (**A**) Schematic of the artificial atom created from Cs with a representation of the attractive confinement potential. (**B**) Constant-current STM image of an artificial atom derived from eight Cs atoms arranged in a ring structure (lateral scale: 5 nm, $\Delta z$ = 300 pm). (**C**) Orbital map of the bound state of the artificial atom in B, obtained at the voltage marked with a red dashed line in D ($V_S$ = -91 mV). The white circles were added to represent the Cs atoms for clarity. (**D**) d$I$/d$V$ spectroscopy measured at the indicated black dot in B, revealing a bound state (red dashed line) within the bulk band gap. (**E**) Constant-current STM image of two artificial atoms separated by $d \approx$ 11 nm (lateral scale: 5 nm, $\Delta z$ = 300 pm). (**F**) d$I$/d$V$ spectroscopy measured sequentially along the dashed white line ($x$) marked in E. The $\sigma/\sigma^*$ indicate the bonding/antibonding orbitals, which are separated by $\Delta E_{12}$ = 46 mV. Insets in F represent spatial distribution of the bonding ($\sigma$) and anti-bonding ($\sigma^*$) orbitals registered at $V_S$ = -133 mV. And $V_S$ = -86 mV, respectively. The white circles were added to represent the positions of the Cs atoms for clarity. (**G**) Distance-dependence of the coupling strength $\Delta E_{12}(d)$ between two artificial atoms as a function of separation $d$. The inset shows a textbook schematic representation of squared wave functions of bonding and antibonding orbitals.

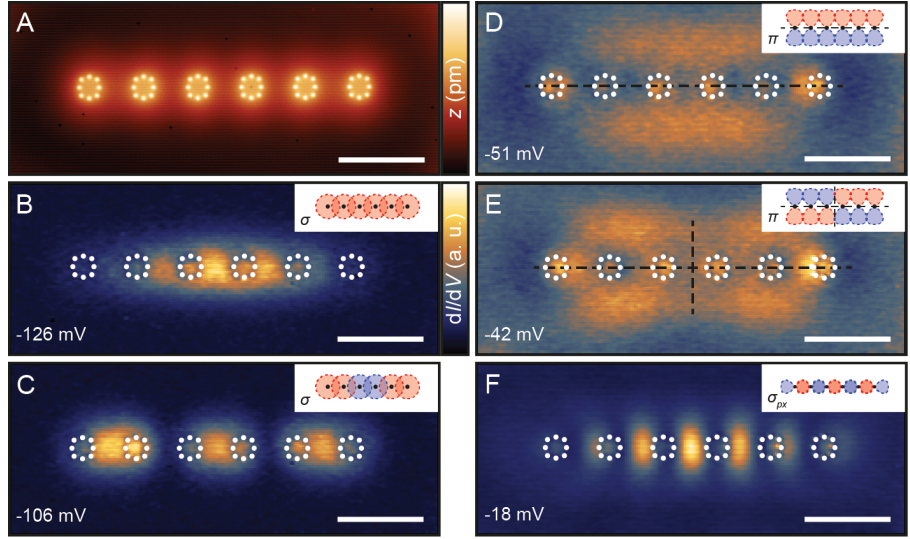

**Fig. 2. Multi-orbital states in an artificial molecular chain.** (**A**) Constant-current STM image of a linear chain composed of six artificial atoms with equal separation $d \approx 12.8$ nm ($V_S = 50$ mV, lateral scale: 20 nm, $\Delta z = 300$ pm). (**B-F**) Orbital maps obtained at the voltages indicated in lower left corner of each image ($z_{offset} = -140$ pm, $V_{mod} = 2$ mV, lateral scale: 20 nm). The white circles represent the Cs atomic positions. A schematic representation of atomic-like orbital contributions is provided in the inset of each map, where red and blue colors represent a positive/negative sign of the depicted wave function, respectively. (**B-C**) Two $\sigma$ orbitals: 1st and 3rd. (**D-E**) Set of two $\pi$ orbitals. Black lines indicate the location of the nodal planes. (**F**) $\sigma$ orbital originating from artificial atomic $p_x$ orbitals.

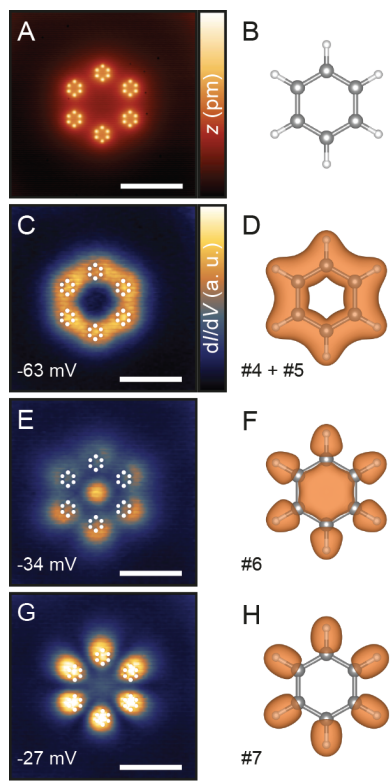

**Fig. 3. Artificial benzene and its orbital structure.** (**A**) Constant-current STM image of six artificial atoms arranged into a benzene structure with separation of $d \approx 10.5$ nm). (lateral scale: 20 nm, $\Delta z = 300$ pm). (**B**) The ball-stick model for the benzene molecule in the DFT calculations. (**C, E, G**) Orbital maps obtained at the voltages indicated in the lower left corner of each image ($z_{\text{offset}} = -140$ pm, $V_{\text{mod}} = 2$ mV, lateral scale: 20 nm). The white circles were added to represent the Cs atoms for clarity. (**D, F, H**) Set of benzene VMOs obtained from the DFT calculations corresponding to orbital maps on the left. The calculations represent the charge density and include the summed charge densities for degenerate orbitals in D. The orbital order number is indicated in the lower left corner of each image.

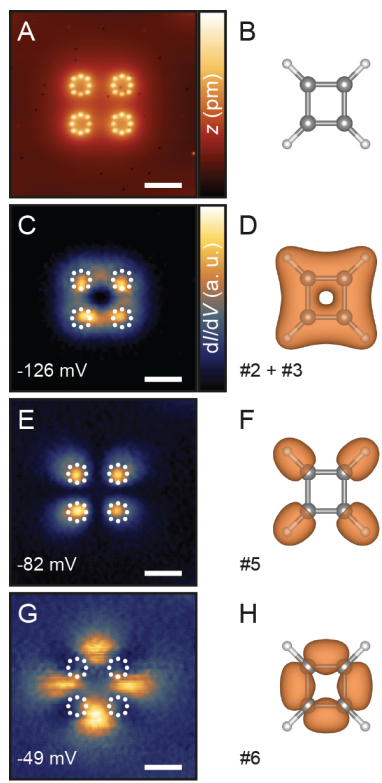

**Fig. 4. Artificial cyclobutadiene and its orbital structure.** (**A**) Constant-current STM image of four artificial atoms arranged in a cyclobutadiene structure with $d \approx 11$ nm (lateral scale: 10 nm, $\Delta z$ = 300 pm). (**B**) The ball-stick model used for the cyclobutadiene molecule in the DFT calculations. (**C, E, G**) Set of orbital maps obtained at voltages indicated in the lower left corner of each image. The white circles were added to represent the Cs atoms for clarity ($z_{\text{offset}}$ = -100 pm, $V_{\text{mod}}$ = 1 mV, lateral scale: 10 nm). (**D, F, H**) Set of cyclobutadiene VMOs obtained from the DFT calculations corresponding to orbital maps on the left. The calculations represent the charge density and include the summed charge densities for degenerate orbitals in D. The orbital order number is indicated in the lower left corner of each image.

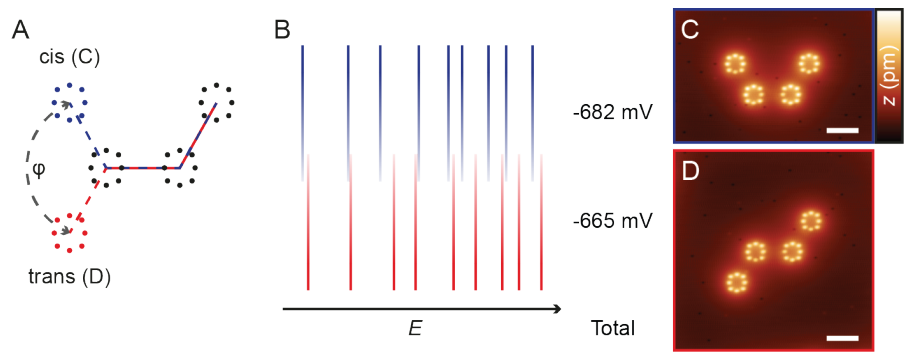

**Fig. 5. Comparison of artificial *cis*- and *trans*-butadiene.** (**A**) Schematic model of the artificial molecular structure built from four artificial atoms in both *cis*- and *trans*-conformations. (**B**) Schematic comparison of the energies of all nine orbitals from the ground state up to $E_F$ of the *cis*- (blue) and *trans*-conformer (red) as identified from STS. The total energy is the sum of these eigenenergies. (**C, D**) Constant-current STM image of (**C**) the *cis*-, and (**D**) the *trans*-conformer (lateral scale: 10 nm, $\Delta z$ = 300 pm).

# Supplementary Materials

**Table of Contents:**

**Materials and Methods**

All experiments were performed under ultra-high vacuum (UHV) conditions with an Omicron low-temperature STM at a base temperature of $T = 4.4$ K. Electrochemically etched W tips cleaned by standard *in situ* procedures were covered with Au by means of dipping and further characterized on a clean Au(111) surface. Undoped InSb crystals were glued to sample holders using a conductive epoxy glue, with an additional stripe of glue along the edge of the crystal to ensure electrical contact to the surface and cleaved in UHV. After the sample was cooled down to the base temperature, Cs was evaporated from a $Cs_2CrO_4$ dispenser (SAES getters) directly onto the cold surface. We used a typical surface coverage of 0.006 ML (ML defined based on adsorption site described below), as determined by counting individual atoms. Unless stated otherwise, STM topography images were acquired in constant-current mode with typical tunneling parameters of $V_S = 200$ mV and $I_t = 20$ pA. Atomic manipulation was performed using lateral manipulation, with various types of tips treated on the surface, with typical tunneling parameters in the range of $V_S = 12 - 20$ mV and $I_t = 50 - 300$ pA.

Each STS spectrum was acquired after stabilizing the tip at $V_S = 50$ mV, $I_t = 100$ pA. Unless stated otherwise, lock-in parameters were set to $V_{mod} = 1$ mV$_{rms}$ and between $f = 667.1 - 941.1$ Hz. Orbital maps were obtained by using a multi-pass method allowing a single line to be scanned multiple times with different parameters for each pass. Taking this approach was necessary to obtain a significant signal without moving the atoms. Here, the data was obtained in line-by-line fashion where initially one line of the structure's contour was obtained in constant-current mode at the standard tunneling parameters ($V_S = 200$ mV, $I_t = 20$ pA), followed by a line with the same lateral and vertical coordinates (i.e. feedback loop turned off) where the $dI/dV$ signal was registered at a particular voltage corresponding to specific energy derived from the STS spectra. Whenever needed, the signal strength was controlled by adding or subtracting a constant $z$ offset. The process was repeated for each line of the image and for all the voltages of interest. All data has been processed using MATLAB and Gwyddion software. All orbital maps are filtered using a mean filter with a 2 px probing window, as well as with standard plane subtraction methods where necessary.

Most parameters for the Schrödinger equation used to calculate the ground and excited bound states of electrons in the 8-Cs ring structure on InSb(110) were obtained using density functional theory (DFT). To this end we constructed a unit cell with an isolated Cs atom on a 4x4 InSb(110) surface supercell consisting of 4.5 layers with 20 Å vacuum spacing and passivated dangling bonds at the bottom. All DFT calculations were performed within the generalized gradient approximation (GGA-PBE) (*49*) using projector augmented wave basis sets as implemented in the Vienna *ab initio* simulation package (VASP) (*50*). In these calculations we set the energy cutoff to 300 eV, the energy convergence criteria to $10^{-6}$ eV, and used 3x3x1 $k$-meshes for the Brillouin zone integrations. The effective mass $m^* \sim 0.02\ m_e$ of the 2D electron gas was taken from experimental results (*33*). The total residing charge $Z \simeq$ +0.8 at the Cs atom was estimated from a charge transfer analysis. This positive charge residing at the Cs position creates an effective potential which is screened by the InSb surface with dielectric constant $\varepsilon = 8.9$ and the 2D electron gas itself, which was considered via a Yukawa screening length $\lambda \sim 10$ nm derived from comparison to experimental data.

The partial charge densities of benzene $C_6H_6$, *trans*- and *cis*-butadiene $C_4H_6$, cyclobutadiene $C_4H_4$, triangulene $C_{22}H_{12}$, and artificial $C_6$ and $C_4$ molecules were obtained from non-spin polarized DFT (GGA-PBE) calculations using VASP package with $C_h$ and $H_h$ type PAW potentials. In these calculations we have set the energy cutoff to 700 eV and the energy convergence criteria to $10^{-6}$ eV. Unit cell for isolated molecules with optimized geometries was chosen to be large enough to avoid spurious interactions with periodic replicas. It results in a qualitative picture of charge density distribution for low-lying molecular states sufficient for comparison with experimental STM data. All structural models and charge distributions obtained from DFT calculations has been visualized using VESTA software. (*51*)

**Supplementary text**

**S1 - Determination of the Cs adsorption site on InSb(110)**

We started by identifying the sublattices of the InSb(110) substrate. According to Whitman *et al.* (*29*), the Sb and In sublattices have an ionic character and could be imaged via STM at $V_S$ = -1.0 V and $V_S$ = 1.2 V, respectively. Fig. S1A and B clearly show that the sub-lattices were shifted with respect to each other by a half unit cell in horizontal direction, as expected. However, our STM images were typically obtained at positive bias voltages close to $E_F$ where the Sb sublattice was imaged (Fig. S1C). Fig. S1D shows that Cs atoms were always adsorbed in hollow sites of the Sb unit cell.

For a theoretical assignment of the Cs adsorption site on the InSb(110) surface, we performed DFT calculations of an isolated Cs atom on the surface of a 4×4 InSb(110) surface supercell using the experimental bulk lattice constant $a$ = 6.479 Å and applied periodic boundary conditions. In $z$ direction (perpendicular to the surface) we used 4.5 InSb layers and a vacuum spacing of 20 Å to reduce artificial interactions between the slabs. Dangling bonds at the bottom of the slab were passivated by pseudo-hydrogen atoms, as described in Ref. (*52*). The exact positions of the Cs adatom and the two uppermost layers of the InSb(110) substrate were optimized. This led to a corrugation of the surface, pushing the Sb atoms up and shifting the In atoms down, in agreement with STM data and previous computational results (*33*). Furthermore, the calculations confirmed that the experimentally identified adsorption site has the lowest energy (Fig. S2).

**S2 - Electronic structure of clean and Cs-covered InSb(110)**

The undoped InSb crystals were intrinsically $n$-doped. STS experiments revealed that the conduction band minimum (CBM) was typically about 50 meV above $E_F$ (Fig. S3A). The exact position of the onset of both valence and conduction bands could be identified more clearly in a semi-log plot. Upon Cs adsorption, a step-like feature with increased d$I$/d$V$ signal intensity at higher voltages became visible in STS, which is a fingerprint of the two-dimensional electron gas (2DEG) with a band onset at around $V_S$ = -90 mV (Fig. S3B).

Bulk InSb is a semiconductor with a narrow band gap of about 235 meV (*28*). Within the slab geometry, the gap depends on the number of InSb(110) layers. On the DFT level it has been shown that 42 layers are needed to reproduce the experimental band gap (*52*), which was beyond the feasibility of our supercell calculations. For a reasonable balance between accuracy and computational feasibility, we used 4.5 layers (one layer marked with black lines in Fig. S2B) of InSb(110) (if not stated otherwise). Cs deposition yielded a charge transfer from the outer Cs $6s^1$ shell to the InSb(110) surface, which effectively dopes the lowest conduction band of InSb(110). We identified this partially occupied lowest band as the origin of the 2DEG (*33*), which has a parabolic dispersion $E(k) = (\hbar k)^2/(2m^*)$ centered at the $\bar{\Gamma}$ point (Fig. S4). The effective mass $m^*$ depended on the doping concentration and varied within the range 0.014 – 0.02 $m_e$ (*33*). Our 4×4 surface supercell with one Cs atom and 4.5 layers of InSb(110) corresponds to a large doping concentration, with effective masses between $m^*_{x(y)}$ ≃ 0.04 $m_e$ (0.08 $m_e$). Note that the effective mass of this band was also affected by the number of involved InSb(110) layers (Fig. S4). Specifically, the difference in effective mass along the *x* vs. *y* direction reduces when the number of InSb substrate layers is increased.

We calculated the transferred charge density $\delta_\rho$ as the difference between the charge densities of the full supercell Cs/InSb(110), the pure substrate InSb(110), and the isolated Cs atom according to:

$$\delta_\rho = \rho_{full} - \rho_{InSb} - \rho_{Cs} \quad \quad \text{S1}$$

From this we found that the transferred charge is mostly confined to the InSb(110) surface near the Cs impurity (Fig. S5). To analyze the total transferred charge from the Cs adatoms to the InSb(110) surface we integrated $\delta_\rho(r)$ outside a sphere of $R_c$ centered on the Cs atom. The result is depicted in Fig. S5 and shows a maximum of about 0.22 electrons far away from the impurity atom. This demonstrates a strong delocalized behavior similar to the 2DEG. Additionally, we performed a Bader charge analysis resulting in an approximate total charge of Cs equals to 0.82, which suggests that 0.18 of the $6s^1$ electron density has been transferred to the InSb surface (Fig. S6). Both values are thus in reasonable agreement with the experimental estimate of about 0.4 given the applied DFT approximations, especially the number of InSb layers, the surface supercell size, and using the PBE functional.

## S3 – States of Cs ring structure

In order to verify that the resultant potential of the Cs ring structure is well approximated by a circularly symmetric $1/r$ potential in the *x-y* plane at sufficient separation from the Cs-ring structure, we calculated its ground and excited states via solving the Schrödinger equation on a disk with radius $R$ using the boundary condition $\Psi_n(r = R, \theta) = 0$:

$$-\frac{1}{2m^*}\Delta\Psi_n(r,\theta) + V(r,\theta)\Psi_n(r,\theta) = E_n\Psi_n(r,\theta) \quad \quad \text{S2}$$

Here we used the effective potential with Yukawa screening parameter $\lambda$, similar to the approach used in Ref. (*33*):

$$V(r) = -\frac{Z}{\varepsilon}\sum_{i}^{8}\frac{1}{|r-r_i|}e^{-\frac{|r-r_i|}{\lambda}} \quad \quad \text{S3}$$

where $r_i$ was the $i^{th}$ Cs atom position in the Cs ring structure as used in the experiments. We set $Z = 0.8$, based on our Bader charge analysis, and the experimental dielectric constant (*53*) was set to a vacuum plus InSb $\varepsilon = (16.8+1)/2$, where the factor of 2 effectively accounted for the surface geometry. The experimental effective mass $m^* = 0.02\ m_e$ was used in our simulations (*33*). The Yukawa screening length $\lambda = 10$ nm was fitted to the experimental data.

In the ground state, we found a radially symmetric charge density with no nodal planes, which mostly resided inside the structure but also contained a significant intensity outside the structure with a decaying tail, which could be identified as a *s* state. In addition, the calculation showed two excited states at 209 meV and 214 meV with $p_x$ and $p_y$ spatial distribution, respectively, extending far outside of the cluster (Fig. S8D). The degeneracy was lifted due to the broken rotational symmetry of Cs ring structure, arising due to the combination of the discreteness and symmetry of the underlying InSb lattice. Overall, the ground state charge density was in good agreement with the experimental data extracted as a maximum intensity of the STS signal at each position (Fig. S7B,C). The STS was performed in a constant height mode with a stabilization point set in the center of the structure, as marked in Fig. S7A.

Fig. S8 presents evolution of calculated charge density for varying parameters $\lambda$, $m^*$ and $Z$. Variations of the screening length $\lambda$ in the Yukawa potential (Eq. S3) modified the shape of the charge density. The charge densities obtained with smaller $\lambda$ were in better agreement with the experimental data inside the structure, however, larger $\lambda$ were required for a better agreement outside the cluster. The presence of significant intensity outside of the structure could be attributed to the poor screening of the surface, owing to the low free carrier density, and was vital to simulate artificial atoms as well as enables their coupling. Larger effective masses $m^*$ as well as larger $Z$ in the Yukawa potential resulted in more localized charge density distributions. All calculated energies of the ground and excited states are given in Table 1. Overall, we found that the calculated results were qualitatively robust to variations of the involved parameters.

## S4 - Examples of different Cs structures

We experimentally tested the robustness of the artificial atom's bound state and its isotropic character by building three other structures representing an artificial atom, in addition to the nearly circular ring structure consisting of eight Cs atoms (8-Cs ring) presented in Fig. 1A-D These structures are presented in Fig. S9: (i) a nearly hexagonal structure consisting of six Cs atoms (6-Cs ring), (ii) a nearly square structure consisting of eight Cs atoms and a (iii) a cross structure consisting of eight Cs atoms. Here, 'nearly' refers to the fact that the Cs atoms could be manipulated only to hollow adsorption sites of the InSb(110) lattice (see Section S1), hence the rectangular surface unit cell (lattice constants: $a_x$ = 4.58 Å and $a_y$ =6.48 Å) only permitted discretized patterns. The representative structural models are displayed in Fig. S9A,D,G,J, and their respective constant-current STM images are presented in Fig. S9B,E,H,K. All structures exhibited a localized resonance state in STS as presented in Fig. S9C,F,I,L. The resonance energy positions for all three of the presented structures based on eight Cs atoms were similar ($V_S \approx$ -90 mV). On the other hand, the bound state of the 6-Cs hexagonal structure was shifted toward the Fermi level, to $V_S$ = -79 mV. It also displayed lower intensity and increased width, presumably due to a slightly increased coupling to the surrounding 2DEG. Overall, from a comparison of many Cs structures on various samples, we found that the energy of the bound state was sensitive to a number of parameters, including the average Cs concentration, subsurface defects, as well as the distance to other Cs atoms.

For all but the square structure we also mapped the spatial distributions of these bound states by acquiring orbital maps. We found that at sufficient lateral distance from the outermost Cs atoms (i.e. much larger than the Cs-Cs distance within each respective structure), the orbital distribution was isotropic. This clearly demonstrated that the symmetry of the Cs structure, including small deviations, did not affect the bound state's *s*-like character in the *xy* plane.

## S5 - Two coupled artificial atoms

STS spectra of two coupled artificial atoms showed two states, as presented in Fig. 1E. The first state was located at $V_1$ = -133 mV with maximum d$I$/d$V$ signal intensity between the artificial atoms. The second state was found at $V_2$ = -86 mV with d$I$/d$V$ signal intensity at the artificial atomic sites and no intensity in between. To confirm that these states can be identified as bonding and antibonding orbitals derived from the LCAO of the artificial atomic *s* orbitals, we acquired orbital maps at the two peak energies, shown in inset of Fig. 1F. These clearly reflected the LDOS distribution expected for *s*-derived bonding (σ) and antibonding (σ*) orbitals with an antinode and a node between the artificial atoms, respectively.

The distance-dependent energies could be reproduced with a simple tight-binding model where the energies of the σ and σ* states are given by $E_{\sigma,\sigma^*}(R) = \varepsilon(R) \pm t(R)$ with $\varepsilon(R) = E_0 + E_1 \exp(-R/R_\varepsilon)$ and the hopping parameter $t(R) = t_1 \exp(-R/R_t)$. We found excellent agreement for the fit parameters $E_0$ = -83 ± 2 meV, $E_1$ = -75 ± 13 meV, $t_1$ = -92 ± 6 meV, $R_\varepsilon$ = 8.3 ± 1.6 nm, and $R_t$ = 8.2 ± 0.4 nm. For example, for a distance of $R$ = 12.824 nm, as used in the six-site chain (see Section S6), the hopping parameter is $t$ = -19 ± 2 meV. For twice the distance, it reduces to $t$ = -4 ± 1 meV, already indicating that next-nearest neighbor coupling in chains is negligibly small.

We also analyzed the bonding peak widths for dimers of 6-Cs and 8-Cs ring structures at various separations *d* (see Fig. S10). The FWHM of each structure was independent of *d*, within the uncertainty of the fit analysis. For the 6-Cs dimers, we observed a FWHM ≃ 8.6 ± 0.7 mV. For the 8-Cs dimer, FWHM ≃ 8.3 ± 0.9 mV. Hence, within the determined error there was no difference in widths of the bound state peaks, and hence they also did not depend on the energy of the bound state.

Bonding-antibonding-like orbitals in coupled artificial structures have been observed on other surfaces with STM-based methods (*25-27, 34*). For those using metallic surfaces, the bonding-antibonding splittings were typically smaller than the width of these states (typically 10-50%). Moreover, to resolve the coupling of such states on metallic surfaces, it has been necessary to subtract the contributions of the underlying metallic bulk, which has typically been a significant fraction of the conduction (*23, 24*). In this work, the states of the artificial atom were in the bulk band gap of the InSb, separated in energy from the bulk bands of the semiconductor, and also below the onset of the surrounding 2DEG. Therefore, no normalization or subtraction was needed, i.e. all differential conductance data provided in this work is d*I*/d*V* raw data. As a result, the bonding-antibonding splitting was much larger than the apparent width of the states themselves, similar to previous studies of In adatoms on InAs(111) (*26, 27*). Compared to previous examples, our data showed the bonding and antibonding peaks the sharpest with FWHM ≃ 8 mV. We also compared widths of the bonding state for dimers at various *d* as well as for dimers with 6 Cs atoms per artificial atomic site, and in all cases found a FWHM = 8.5 ± 0.9 mV (see Fig. S10).

Finally, we note that the underlying conduction band of InSb was unpinned, in contrast to the example of In/InAs(111)-A, where the conduction band is pinned due to the intrinsic surface state with fixed carrier density (*26*).

## S6 - Linear chain composed of six artificial atoms

As discussed in the main text, it was necessary to show that the coupling of artificial atomic states also leads to higher symmetry orbitals. As discussed above (Fig. S8D), the separation in energy between *s* and *p* orbitals was most likely too large to observe in the d*I*/d*V* spectra of single artificial atoms and dimers, because the π orbitals would reside in the conduction band of the InSb. Figure S11 shows a constant-current STM image and STS spectra of the same linear chain composed of six artificial atoms (8-Cs ring structure) presented in Fig. 2A. To identify resonances, we measured STS at each point along a scan line above the structure for two distances, as shown by the dashed line in Fig. S11A. Each point spectra were stabilized individually with the same stabilization parameters ($V_S$ = 50 mV, $I_t$ = 100 pA). In Fig. S11C, the resulting d$I$/d$V$ signal intensity is plotted in a false-color scale as a function of position and voltage (like the plot presented for the dimer in Fig. 1F). In order to also identify resonances that have increased LDOS further away from the central axis of the chain, as to be expected for π orbitals, we repeated the measurement along a scanline that is further displaced in $y$ direction, as shown in Fig. S11B. The resulting set of STS is shown in Fig. S11D. A wide range of resonances could be identified in these spatially resolved STS and subsequently imaged as an orbital map (Fig. S12). The first hierarchy of six states found between $V_S$ = -126 mV and -66 mV (Fig. S12A-F) was a series of σ molecular orbitals derived from a superposition of the artificial atomic $s$ orbitals. The ground state (Fig. S12A) was a superposition with the same sign of all artificial atomic wave functions (see inset schematic in Fig. 2B), i.e. they all coupled in a bonding fashion. Successively, an increasing number of nodes was found, i.e., the second orbital (Fig. S12B) contained one node in the center of the chain, the third state (Fig. S12C) contained two nodes, the fourth (Fig. S12D) contained three nodes, the fifth state (Fig. S12E) contained four nodes and the sixth state (Fig. S12F) contained five nodes. The spatial distribution of orbitals confirmed that these resonances could be identified as σ molecular orbitals derived from a LCAO of $s$ orbitals of the artificial atoms. Their energies were reproduced well using a tight-binding model with only nearest-neighbor hopping and including an on-site energy shift from nearest neighbor sites (*54*), assuming a hopping parameter $t$ = 18 meV, i.e., within the error bar of the analysis from the distance dependence of dimers (see section S5). This confirmed that no next-nearest neighbor coupling needed to be accounted for.

In the range $V_S$ = -60 mV to $V_S$ = -30 mV, the spectra in Fig. S11C shows an apparent gap when measured near the chain. But looking at spectra obtained at a further distance orthogonal to the chain axis (Fig. S11D), this region contains three resonances. Orbital maps at these three energies ($V_S$ = -51 mV, $V_S$ = -42 mV and $V_S$ = -30 mV) are shown in Fig. S12G-I We have identified them as the first three π molecular orbitals derived from a LCAO of $p_y$ orbitals of the artificial atoms. Fig. S12G (also Fig. 2D) shows a fully bonding-type coupling of artificial atomic $p_y$ states, leading to an artificial molecular π orbital with a node plane along the chain axis (see horizontal dashed black line in Fig. 2D) and increasing local density of states (LDOS) above and below the chain. The next π orbital (Fig. S12H and Fig. 2E) contained a faint node in the middle of the chain (indicated by the vertical dashed line in Fig. 2E), which can be rationalized by a change of the phase of the artificial atomic $p_y$ wave functions (see inset in Fig. 2E). This was followed by a third π orbital with two nodes (Fig. S12I). While there should still be three more $p_y$-derived molecular orbitals at higher sample biases, we did not observe them, again presumably since those were located at energies where hybridization with the conduction band may weaken and broaden their LDOS. However, we found another σ orbital at $V_s$ = -18 mV (Fig. 2F) which perfectly resembled the LCAO of artificial atomic $p_x$ orbitals with bonding character, leading to increased LDOS along the chain axis between each of the artificial atomic sites. Fig. S11 shows that the d$I$/d$V$ signal intensity showed additional dispersive behavior at even larger voltages, with distinct LDOS distributions. However, in this energy range the spectral features are washed out, presumably due to an increase of coupling to the 2DEG as well as incipient coupling to the bulk conduction band with increasing voltage. We note that in almost all orbital maps the LDOS intensity (i) showed slight deviations from perfect symmetry, and (ii) was reduced at the ends of the linear artificial molecule. Regarding the former, this may be a minor effect of bulk dopants or the distribution of surrounding Cs atoms. The latter is in accordance with expectations from molecular orbital theory (e.g. Hückel), where the coefficients for each atomic wave function in LCAO are modulated by a particle-in-a-box-like envelope function, reducing the total energy (referred to as delocalization energy) (*55*).

## S7 - Artificial benzene structure and DFT results of benzene molecule

Fig. S13 presents the artificial benzene molecule consisting of six artificial atoms, where each artificial atom was 6-Cs hexagonal structure. We note that owing to the discretization of the possible positions of Cs atoms on the InSb(110) lattice (see Section S4), there was a slight deviation from the perfect hexagonal symmetry. However, the difference in bond lengths of less than 0.2 Å was negligible compared to the spatial extension of the artificial atomic wave functions (see Fig. S7C and Fig. S8A-C). The structural is presented in Fig. S14.

Fig. S15A shows an STM image of the artificial benzene structure, including four points and a dashed line that mark positions where we measured STS (shown in Fig. S15B,C) to identify resonances for orbital maps. Fig. S16 presents orbital maps at all voltages below the Fermi level where a peak was observed in one of the STS spectra. Most of these maps are the same as already presented in Fig. S13 (excerpt is also presented in Fig. 3 of the main text), where they are compared with the first eight calculated valence molecular orbitals (VMOs) of the actual chemical structure of benzene, based on DFT. In the following, we will thoroughly describe and discuss those orbitals.

The lowest state at $V_s$ = -135 mV (Fig. S13C) was an orbital delocalized over the entire hexagonal structure, exhibiting no nodes and maximum LDOS intensity in the center of the hexagon. This is in very good agreement with the characteristics of the charge density of the calculated first benzene VMO (Fig. S13D) at scaled distances. The next state at $V_s$ = -103 mV exhibited maximal LDOS delocalized across a ring that connects all the artificial sites. In comparison to the calculations, the observed intensity was reminiscent of the superposition of the second and third VMOs of benzene (Fig. S13F), which are degenerate. The following orbital map (Fig. S13G) had a similar geometry to the previous map (Fig. S13E), but the intensity was more delocalized toward the circumference of the artificial atomic sites, exhibiting a pronounced hexagonal symmetry. When comparing these calculations, this can be explained by a superposition of the degenerate fourth and fifth benzene VMO (Fig. S13H). The map at $V_s$ = -34 mV (Fig. S13I) displayed a more nodal pattern of six lobes with intensity toward the exterior of the artificial sites and a pronounced lobe in the center of the structure. This is different in the next map taken at $V_s$ = -27 mV, where the lobes at each artificial atom were azimuthally more localized with node lines in between each site, while there was no LDOS at the center of the hexagon. Both the orbital maps in Fig. S13I and S13K agree with the sixth and seventh calculated VMO (Fig. S13J and S13L), respectively. Finally, Fig. S13M shows a map where LDOS was located between each of the artificial atoms, while nodes were present at each of them. This map is in partial agreement with the calculated eighth benzene VMO (Fig. S13N). There was also intensity in the center of the structure, which was not seen in the calculation. The overall qualitative agreement between the orbital maps and the calculated molecular orbitals provides evidence that the artificial atomic $s$, $p_x$, and $p_y$ orbitals have undergone $sp^2$ hybridization to form artificial molecular orbitals.

In addition, we also found a faint resonance at about $V_S$ = -51 mV localized in the center of the artificial benzene ring. This map could not be assigned to a particular valence molecular orbital (VMO) of benzene. We speculate that this may be a signature of the underlying 2DEG. As already discussed for the linear six-site chain in the previous section, additional features in the LDOS were found above the Fermi level (see Fig. S15C). We expect those states to couple to the 2DEG and the bulk conduction band. For this reason, we refrain from discussing the maps above the Fermi level in terms of artificial molecular orbitals.

For more detailed comparison of the experimental orbital maps of the artificial structure with the actual benzene molecule, we show all DFT-calculated occupied benzene VMOs in Fig. S16. Note that degenerate VMOs are presented separately, while superpositions were presented in Fig. 3 and Fig. S13, as the latter is what is expected to be visible in the experiments.

## S8 - Artificial cyclobutadiene structure

Fig. S18 illustrates the artificial cyclobutadiene molecule consisting of four artificial atoms, where each artificial atom was an 8-Cs ring structure. The structural is presented in Fig. S19. In order to identify the energies of interest where a molecular resonance state might be, we performed STS along various lines and points of the structure, some of which are presented in Fig. S20. Fig. S21 presents orbital maps at all voltages below the Fermi level where a peak was observed in one of the STS. Most of these maps are the same as already presented in Fig. 4.

The lowest energy state at $V_s$ = -176 mV (Fig. S18C) showed LDOS centered in the square and delocalized over all four artificial atoms, in agreement with the first cyclobutadiene VMO (Fig. S18D). The second state showed a ring of LDOS with no intensity in the center of the square (Fig. S18E), also in agreement with a superposition of the degenerate second and third VMO (Fig. S18F). The third state at $V_s$ = -83 mV (Fig. S18G) showed LDOS located at the center of each artificial atom, but with intensity extending outward at 45 degrees with respect to the horizontal and vertical axes of the structure. This state could be identified as the fifth cyclobutadiene VMO (Fig. S18H) of the DFT calculation. At $V_s$ = -67 mV, we observed LDOS inside as well as a faint ring outside of the square structure (Fig. S18I). While this orbital map did not agree well with the cyclobutadiene VMO #4 as shown in Fig. S18J, we note that the four outer lobes had the same sign, which is why they fused into a square-shaped ring when imaging the calculated VMO using smaller isosurface values. Still, the much smaller intensity of the ring vs. the central lobe in the experimental map could not be explained this way. We speculate that this effect could be caused by the orbital coupling to the 2DEG, which would delocalize the density of states. Finally, the map at $V_s$ = -49 mV (Fig. S18K) showed LDOS concentrated in between the artificial atoms, in good agreement with VMO #6 (Fig. S18L). We note that this orbital map exhibited a small asymmetry, presumably due to the fact that the structure deviates from a perfect fourfold symmetry (see Fig. S19), but with no noticeable anisotropy on the orbital levels (Fig. S20). The last two orbital maps similarly showed that the LDOS extended laterally more outside the structure than at other energies, further supporting our hypothesis that artificial molecular orbitals above the 2DEG onset coupled to it. Our simulator allows us to explore the interplay of the resultant electronic structure of analogous 2D structures, with and/or without such distortions.

In addition, we also found maps at $V_S$ = -40 mV, $V_S$ = -24 mV, $V_S$ = -15 mV and $V_S$ = -5 mV. These maps could not be clearly assigned to any particular VMO of cyclobutadiene. We speculate that this may be caused by the underlying 2DEG, similar to the case of one orbital of the benzene structure in the previous section. For this reason, we refrain from discussing these maps in terms of artificial molecular orbitals.

For a more detailed comparison of the experimental orbital maps of the artificial structure with the actual cyclobutadiene molecule, we show all occupied cyclobutadiene molecular orbitals calculated by non-spin polarized DFT in Fig. S20. In this calculation, the positions of atoms were optimized and the resulting structure had $C_4$ symmetry. The degenerate orbitals are presented separately, while superpositions were presented in Fig. 4 and Fig. S18, as the latter is what is expected to be visible in the experiments.

### S9 - Artificial *cis*-butadiene structure

Fig. S23 illustrates the structural model of the artificial *cis*-butadiene molecule consisting of four artificial atoms at ~120° angles and with a distance of $d \approx 11$ nm from each other, where each artificial atom was 8-Cs ring structure. In order to identify the energies of interest where a molecular resonance state might be, we performed STS along various lines and points of the structure, some of which are presented in Fig. S24. Fig. S25 presents orbital maps at all voltages below the Fermi level where a peak was observed in STS. This data was used to create the diagram in Fig. 5B.

The first four orbital maps ($V_s = -160$ mV, $V_s = -130$ mV, $V_s = -107$ mV and $V_s = -85$ mV) are in good agreement with the first four VMOs of *cis*-butadiene. The next two maps ($V_s = -66$ mV and $V_s = -57$ mV) is comparable to the fifth VMO. The next three maps ($V_s = -40$, $V_s = -27$ mV and $V_s = -10$ mV) can be roughly assigned to the sixth, seventh and eighth DFT-calculated VMO of *cis*-butadiene, respectively. However, each of these orbital maps also contains faint features from orbitals directly above and below the assigned orbital. This is a consequence of energetic overlap of these states, presumably due to the increased hybridization with the 2DEG (see also Section S10 and S12). For a more detailed comparison of the experimental orbital maps of the artificial structure with the actual *cis*-butadiene molecule, we show calculated by non-spin polarized DFT all occupied *cis*-butadiene molecular orbitals in Fig. S26. In this calculation the positions of atoms were optimized, the resulting geometry of molecule is fully planar.

### S10 - Artificial *trans*-butadiene structure

Fig. S27 illustrates the structural model of the artificial *trans*-butadiene molecule consisting of four artificial atoms, where each artificial atom is an 8-Cs ring structure. In order to identify the energies of interest where a molecular resonance state might be, we performed STS along various lines and points of the structure, some of which are presented in Fig. S28. Fig. S29 presents orbital maps at all voltages below the Fermi level where a peak was observed in STS. This data was used to create the diagram in Fig. 5B.

The first four orbital maps ($V_s = -158$ mV, $V_s = -129$ mV, $V_s = -102$ mV and $V_s = -89$ mV) are in good agreement with the first four VMOs of *trans*-butadiene. The combination of the next two maps ($V_s = -61$ mV and $V_s = -50$ mV) is comparable to the fifth VMO while the next three maps ($V_s = -39$, $V_s = -25$ mV and $V_s = -12$ mV) are predominantly assigned to the sixth, seventh and eighth DFT-calculated VMO of *trans*-butadiene, respectively. Again, those orbital maps still contain features from neighboring orbitals due to energetic overlap of the states in this energy range (see also Section S9 and S12). For a more detailed comparison of the experimental orbital maps of the artificial structure with the actual *trans*-butadiene molecule, we show all DFT-calculated occupied *trans*-butadiene VMOs in Fig. S30.

## S11 – Comparison of cyclo-, *cis*- and *trans*-butadiene artificial structures based on 6- and 8-Cs artificial atoms

Fig. S31 illustrates a comparison of all butadiene structures considered in this manuscript, i.e. cyclo- (Fig. S31A-D), *cis*- (Fig. S31E-H) and *trans*- (Fig. S31I-L) conformations, where each was built using two different artificial atom structures, 6-Cs and 8-Cs (see Fig. S9), while the distances between the artificial atoms' centers were kept the same. Fig. S31A,C,E,G,I,K are STM topography images while Fig. S31B,D,F,H,J,L are excerpts of corresponding orbital maps.

Comparing the cyclobutadiene structures, there were two main differences between their electronic structures: energy positions and spatial extents of some of the orbital maps were different. The energy positions of orbital maps with comparable appearance i.e. the lowest energy state ($V_s$ = -138 mV and $V_s$ = -179 mV for 6-Cs and 8-Cs structures, respectively), the second ($V_s$ = -88 mV and $V_s$ = -126 mV) and the fifth state ($V_s$ = -24 mV and $V_s$ = -49 mV) were lower for the 8-Cs cyclobutadiene structure. The orbital maps of the second energy state were comparable in terms of appearance but differed slightly in registered intensity. The orbital maps of the third state of 6-Cs based cyclobutadiene ($V_s$ = -52 mV) had a spatial extent comparable with the fourth state of the 8-Cs structure ($V_s$ = -67 mV), but their ordering with respect to the other orbital maps is different. For 6-Cs based cyclobutadiene we found an additional orbital map at $V_s$ = -37 mV, for which no corresponding orbital map could be assigned for the 8-Cs based structure. In turn, for the 8-Cs based cyclobutadiene, there was an additional map at $V_s$ = -82 mV that could not be found for the 6-Cs structure. Those two maps had different symmetries and hence cannot be considered a reflection of the same molecular orbital. We note that, as discussed in the main text, these limitations in assigning maps to VMOs is presumably due to coupling of the artificial molecular orbitals with the 2DEG at larger energies.

For *cis*- and *trans*-butadiene the main differences were in energy positions of registered electronic states. As in the case of cyclobutadiene, the energies at which given orbital maps appeared were always lower for structures based on 8-Cs artificial atoms.

Hence, in all artificial molecules that were built, the energy levels of all orbitals were always pushed down for the 8-Cs based structures relative to the 6-Cs based structures. Consequentially, we also observed more states below $E_F$. To estimate how many electrons were effectively bound in each of the artificial molecular structures, we counted all states below $E_F$ and assumed that they are doubly occupied. In cases where the comparison to the calculations lead us to conclude that a measured state is degenerate, we counted it twice. When normalizing the resulting number of bound electrons by the number of Cs atoms used for the respective artificial molecule, we found that, irrespective of the number of artificial atoms and the conformation, there were always about 0.4 to 0.5 bound electrons per Cs atom. Interestingly, this value is in accordance with the effective charge donation per Cs atom into the 2DEG (see Section S2), opening up the possibility of controlled "gating".

## S12 – Artificial triangulene structure and DFT results of triangulene molecule

Fig. S32A presents a triangulene-like structure, whose chemical analog is difficult to synthesize. No Kekulé structure exists for triangulene (*46*), and as a consequence, despite having 4n+2 electrons, it turns out to be a biradical (i.e. two singly occupied VMOs), which is extremely reactive (*41, 42*). Only recently, the on-surface synthesis of unsubstituted monomers, dimers and chains has been achieved (*43-45*). Using artificial benzene as a basis, we constructed a structure inspired by triangulene, which we refer to as artificial triangulene in the subsequent discussion. This structure consists of 22 artificial atoms, where each artificial atom was a 6-Cs hexagonal structure, assembled in six fused benzene rings in a form of a triangle (see Fig. S33 for a structural model).

In order to identify energies of interest where a molecular resonance state might be, we performed STS along various lines and points of the structure, some of which are presented in Fig. S34. Due to the large size of this structure and higher number of artificial atoms, the number of resonances within the probed energy window strongly increased compared to the smaller artificial structures. We note that at these small voltage separations, some resonances also overlapped within their widths (FWHM ≈ 10-12 mV). Therefore, it was not possible to acquire orbital maps at discrete energy separations due to the overlapping resonances, but the orbital maps rather represent linear combinations. Consequently, it becomes more difficult to compare the experimental maps with the calculated VMOs of the triangulene molecule.

An excerpt of certain artificial molecular orbitals together with a direct comparison with the calculated VMOs (or their combination) for triangulene is presented in Fig. S32, while we present a full set of observed orbital maps in Fig. S35 and calculated by non-spin polarized DFT a set of VMOs in Fig. S36. In this calculation the positions of atoms were optimized. We found a very good agreement between the first five orbital maps ($V_S$ = -179 mV to $V_S$ = -100 mV) and the first twelve of the corresponding calculated VMOs, as shown in Fig. S32C-L. The further evolution of states at higher energies is visible in both the STS (Fig. S34) and orbital maps (Fig. S35).

Further comparison with the additional orbital maps in Fig. S35 is possible but given the width of each state, maps at a particular voltage pick up superpositions of an increasing number of orbitals, making an unambiguous assignment to states more difficult. The maps at $V_S$ = -96 mV and $V_s$ = -92 mV showed features that are reminiscent of the degenerate VMOs #13-15. The map at $V_S$ = -83 mV showed only weak d$I$/d$V$ intensity, but it resembles the VMO #16. The map at $V_S$ = -74 mV displayed a noticeable roughly triangular shape with an intensity at the center artificial atom of the structure. An LDOS distribution with the same features can also be seen in VMO #17, while the outer intensity is captured in VMOs #18-19. The map at $V_S$ = -62 mV featured strong intensities located within outermost hexagonal rings. This feature can be partially explained by the VMOs #20-21.

For a comparison to the calculations, the maps from $V_S$ = -55 mV to $V_S$ = -25 mV might be considered as a gradual transition between characteristic features found in the VMOs #22-27. The resonances were so close in energy that disentangling the characteristic features of each and assigning a specific VMO was not possible. Finally, the maps at $V_S$ = -18 mV and $V_s$ = -8 mV show a central triangular feature. At first glance, this map seems reminiscent of VMO #28. However, this is the first $p_z$-derived π VMO of triangulene, which we believe cannot be emulated with our quantum simulator platform, owing to its intrinsic 2D character. We speculate that this map could be a consequence of increasing interaction between the artificial orbitals and the 2DEG with increasing energy, which can eventually lead to hybrid orbitals.

**S13 - Comparison of calculated orbitals for molecules with and without H termination**

Figures S37 and S38 present a comparison between two hexagonal and two square molecules. For each comparison one of the molecules contains hydrogen terminations while the other does not. The molecules with hydrogen terminations are the same as presented before, i.e. benzene ($C_6H_6$) and cyclobutadiene ($C_4H_4$). For each molecule, we present the first ten VMOs. We note that for both of these comparisons the shapes of the VMOs were comparable, but their energy ordering changed.

For the $C_6H_6$ to $C_6$ comparison (Fig. S37) there is an agreement for the first five VMOs. The shape of VMO #6 of $C_6H_6$ is similar to that of VMO #7 of $C_6$. At the same time, VMO #6 of the $C_6$ can be compared to VMO #8 of the $C_6H_6$. We note that in this comparison an equivalent of the $C_6H_6$ VMO #7 cannot be found. For the $C_6$ molecule, the VMOs #8-10 are π-orbitals and equivalents of those for $C_6H_6$ can be found at higher energies than the ones presented here.

For the case of $C_4H_4$ to $C_4$ comparison (Fig. S38) there is an agreement for the first three and the last three VMOs. The shape of the $C_4H_4$ VMO #4 can be compared to the $C_4$ VMO #6. The $C_4H_4$ VMO #5 is similar to the $C_4$ VMO #7. The $C_4H_4$ VMO #6 resembles the $C_4$ VMO #4. Finally, the out-of-plane π-VMOs can be found as #7 and #5 for the $C_4H_4$ and $C_4$ molecule, respectively.

## S14 – Further discussion

The simple LCAO picture applicable to the dimers and linear chains (Fig. 1E-G, 2 and S12) as well as the resemblance of the orbital maps for the various artificial 2D molecular structures (Fig. 3-4 and Fig. S13, S16, S18, S21, S25, S29, S3, S3) to VMOs of actual organic molecules is uncanny. They provide evidence that the underlying radially isotropic potential of each artificial atom created not only $s$ but also $p_x$ and $p_y$ orbitals that couple into molecular orbitals. Moreover, we found evidence for carbon-like $sp^2$ hybridization in all structures with 120° angles between artificial atoms. Such bond angles make this type of hybridization more favorable. However, the hybridization is not necessarily driven by geometry alone. This was most obvious for the artificial cyclobutadiene structure, which exhibited a fourfold symmetry but still provided evidence for $sp^2$ hybridization by comparison with the calculated VMOs of actual cyclobutadiene. This is different from previous studies based on confining surface states with particular orbital band character, where the geometry of the artificial structure determined the spatial distributions and energies of the band modulations that mimicked the $s$-, $p_x$- and $p_y$-like states (*47, 56*).

Despite the astounding similarity of the orbital maps to analogue molecules, it is important to be careful not to assign a particular element (e.g. carbon) to the artificial atoms in the given structure. Not only are the artificial atoms of 2D nature, but it is also unclear how many electrons were bound to the artificial atoms and molecules. For example, in the case of artificial benzene, we identified six different states below $E_F$, of which two were doubly degenerate. Hence, about 16 electrons should have been trapped within the structure, which corresponds to about 2.7 electrons per artificial atom (or about 0.44 electrons per Cs atom). In comparison, each carbon atom in benzene provides four electrons from the 2$s$ and 2$p$ levels to form VMOs. In a real molecule, the chemical (and hence electronic) structure is determined by this electron occupation of the chosen elements, and the exact splitting between the $s$, $p$ etc. states. In our artificial molecular structures, we still achieved carbon-like $sp^2$ hybridization, despite a different filling, by fixing the geometry. As of now, it remains unclear why artificial cyclobutadiene orbitals also showed this hybridization.

Nevertheless, we were able to change the electron occupation, i.e. to effectively "gate" the artificial structure, by varying the number of Cs atoms used per artificial atomic site. While this was not yet obvious from spectra taken on individual sites (Fig. S9), a comparison of d$I$/d$V$ spectra on artificial butadiene made from eight Cs atoms per site (8-Cs structure) with those of the same structure built from six Cs atoms per site (6-Cs structure) shows a clear difference (Fig. S31). In the 6-Cs structure, all states were pushed higher in energy, and overall, less filled states were found below the Fermi energy, resulting in a lower number of bound electrons in the artificial molecule. We found that this trend occurred throughout all artificial molecules where we compared 6-Cs and 8-Cs structures with each other (see Section S11). Our estimates lead to an effective electron occupation of the artificial molecules ranging from 0.4 to 0.5 electrons per Cs atom, for both 6-Cs and 8-Cs structures and independent of the exact molecular structure. We note that assigning integer charges is a simplification, because of weak coupling to the 2DEG for all states located between the 2DEG onset energy and $E_F$, which also manifested in larger widths of those states in d$I$/d$V$ spectra. This not only posed an uncertainty in determining exactly how many orbitals were located below $E_F$ in some cases, but this also can lead to non-integer charges, reflecting that a many-body picture may be a more accurate description.

It is thus surprising that we find such a good agreement with VMOs, considering that (i) our quantum simulator is intrinsically 2D in nature, (ii) the geometry of our artificial structures is fixed and cannot relax, and (iii) we did not emulate the presence of additional artificial sites that would play the role of the hydrogen atoms in the artificial molecule. Concerning the latter point, we also calculated the electronic structure of a $C_6$ ring as well as a $C_4$ ring without H (see Fig. S37 and S38). While the spatial distributions of the calculated $C_6$ VMOs were found to be qualitatively identical to those of benzene, their energetic order is only the same for the first five VMOs but deviates from both benzene and the observations of our artificial structure (Fig. 3 and Fig. S13) for higher VMOs. At the moment, it is unclear why our artificial molecular orbitals are in better agreement with $C_6H_6$ and $C_4H_4$ rather than $C_6$ and $C_4$, respectively. This point is particularly important for cyclobutadiene, where the $sp^2$ hybridization, despite the angular distortion, is stabilized by the hydrogen atoms bound to each carbon. The lack of additional artificial atoms that would take over the role of hydrogen in our structures might have also led to unhybridized orbitals (as we observed for the linear structure in Fig. 2), as the overlap of unhybridized $p_x$- and $p_y$-orbitals would also be large. More systematic studies that include varying bond angles and distances as well as coupling to the 2DEG may shed more light on how hybridization can be influenced.

Based on this platform, more well defined multi-orbital states can be defined and hybridized, which are strongly decoupled from bulk bands. Our observations open exciting perspectives toward tuning the frontier orbitals of artificial molecular structures. For instance, we may be able to push the highest occupied molecular orbital (HOMO) across $E_F$, which may lead to a singly occupied state (SOMO). This may induce a spin moment in the artificial molecule. Besides, our platform can enable quantitative comparisons on how the electron occupation (i.e. the charge state) influences the frontier orbital densities, independent from relaxation effects(*57, 58*). Furthermore, it will be interesting to explore heterostructures in future experiments, i.e., using artificial atoms with varying numbers of Cs atoms per site. The roles of the triangular confinement potential, the 2DEG, and their interplay with the simulated electronic structure still need to be better understood. For example, due to the surface, only lower dimensional orbital structure could be simulated, limiting analogues to 3D orbital structures based on $p_z$ symmetry. Moreover, the electronic structure of the artificial atom must be better understood.

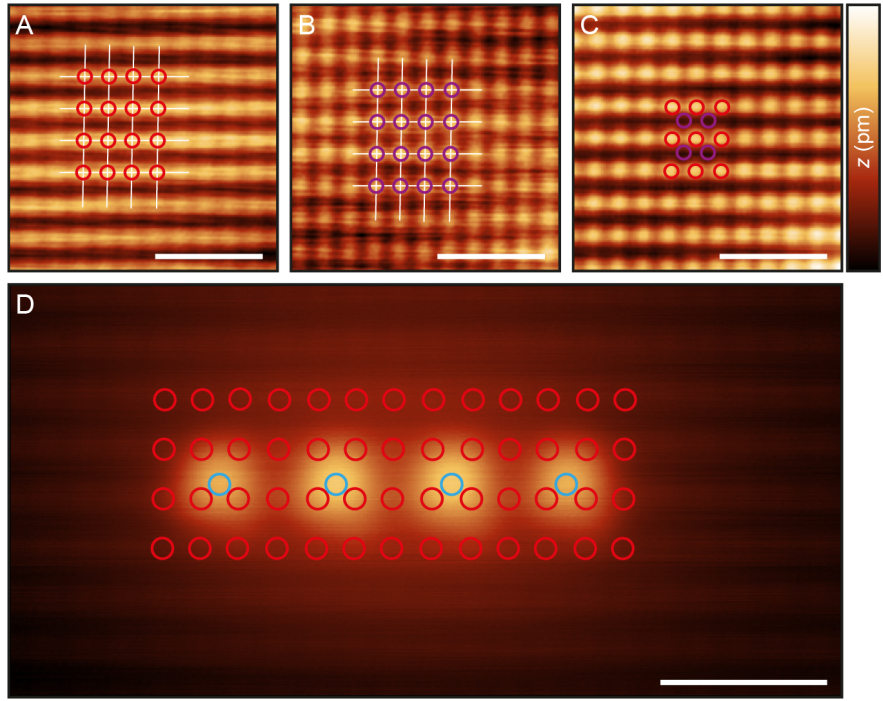

**Fig. S1. Adsorption site of a Cs atom.**
(**A**) Constant-current STM image of Sb sublattice obtained at $V_S$ = -1 V (lateral scale: 2 nm, $\Delta z$ = 20 pm). (**B**) Constant-current STM image of In sublattice obtained at $V_S$ = 1.2 V (lateral scale: 2 nm, $\Delta z$ = 300 pm). (**C**) Constant-current STM image of Sb sublattice obtained at $V_S$ = 50 mV (lateral scale: 2 nm, $\Delta z$ = 25 pm). (**D**) Constant-current STM image of Cs atoms adsorbed on InSb surface obtained at $V_S$ = 50 mV (lateral scale: 2 nm, $\Delta z$ = 300 pm). Red, purple and cyan circles mark positions of Sb, In and Cs atoms, respectively.

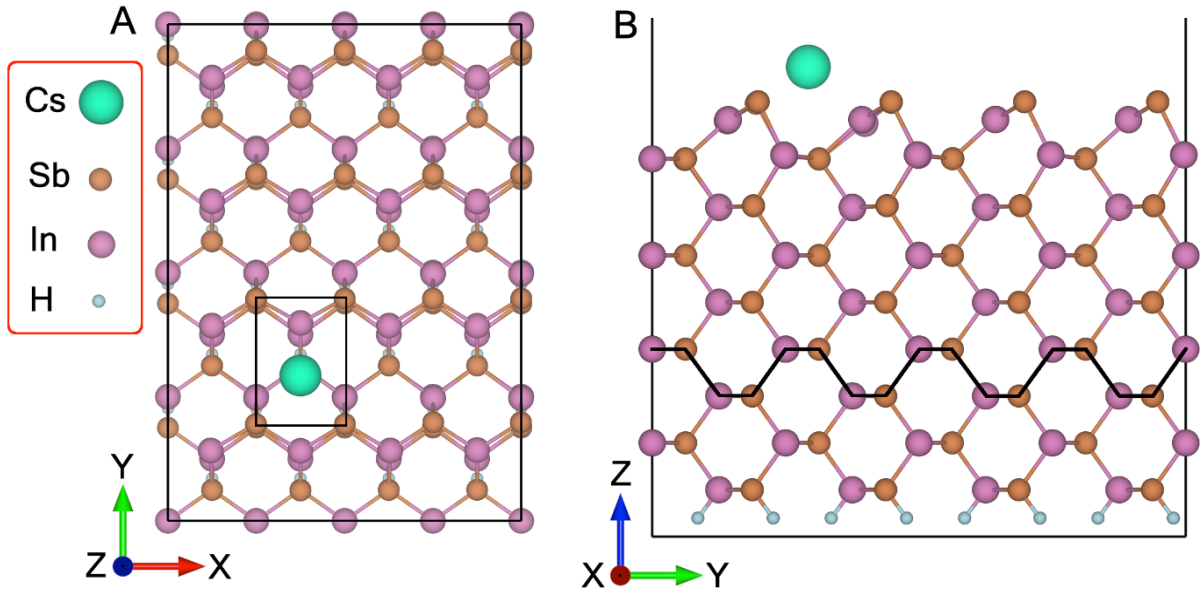

**Fig. S2. Supercell of InSb(110) substrate with single Cs atom.**
(**A**) Top and (**B**) side views of 4x4 InSb(110) supercell. X, Y and Z axes correspond to [1$\bar{1}$0], [00$\bar{1}$] and [110] crystallographic directions, respectively. The monolayer of InSb(110) is schematically marked as a black line.

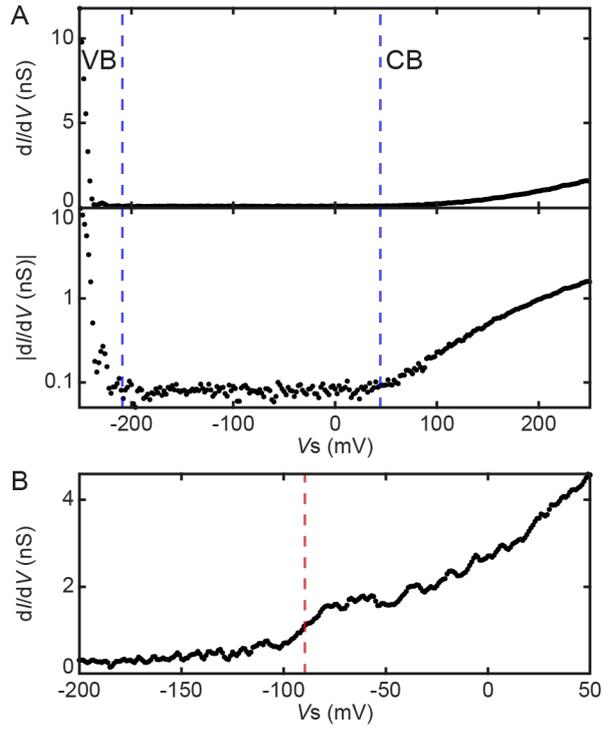

**Fig. S3. STS of bare InSb(110) and the Cs-covered surface.**
(**A**) STS of clean undoped InSb(110) surface plotted with two scales: (top) linear and (bottom) semi-log. Blue dashed lines mark valence band maximum (VBM) and conduction band minimum (CBM). (**B**) STS of Cs-covered InSb(110) surface with a characteristic step-like feature of the 2DEG, whose onset is marked by a red dashed line.

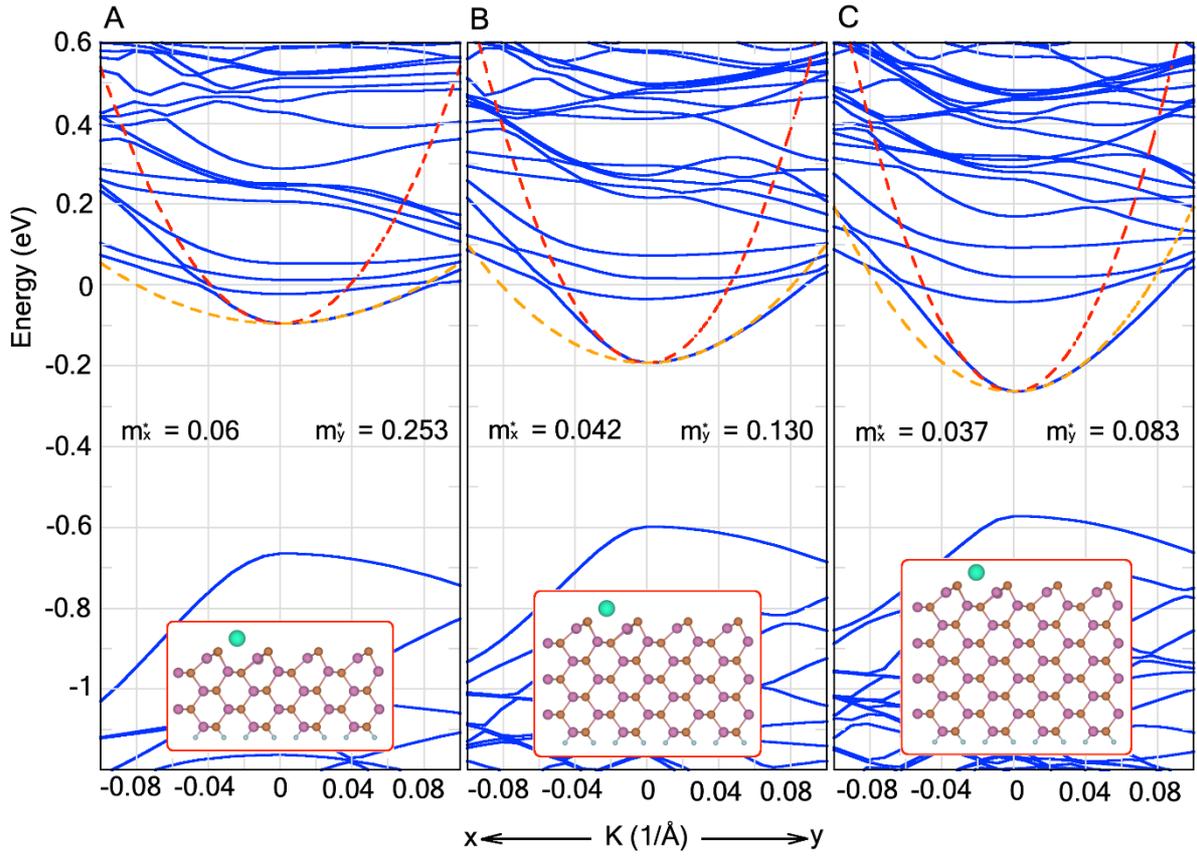

**Fig. S4. Band structure of 4×4 Cs/InSb(110).**
Evolution of the bands with different number of InSb layers (shown as inserts). Parabolic dispersions were fitted along the *x*-direction (red dashed lines) and *y*-direction (orange dashed lines), respectively, by equation $E(k) = (\hbar k)^2/(2m^*)$.

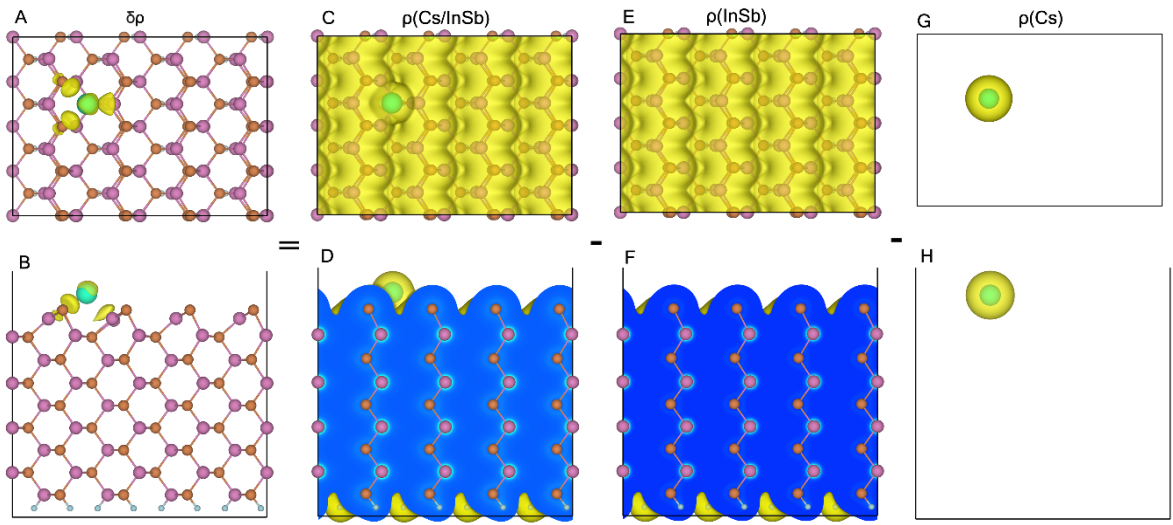

**Fig. S5. Charge transfer from Cs impurity atom to InSb(110) surface.**
Resulting transferred charge (A-B) corresponds to difference between the charge density of the full supercell Cs/InSb (C-D), the pure substrate InSb (E-F), and the isolated Cs atom (G-H).

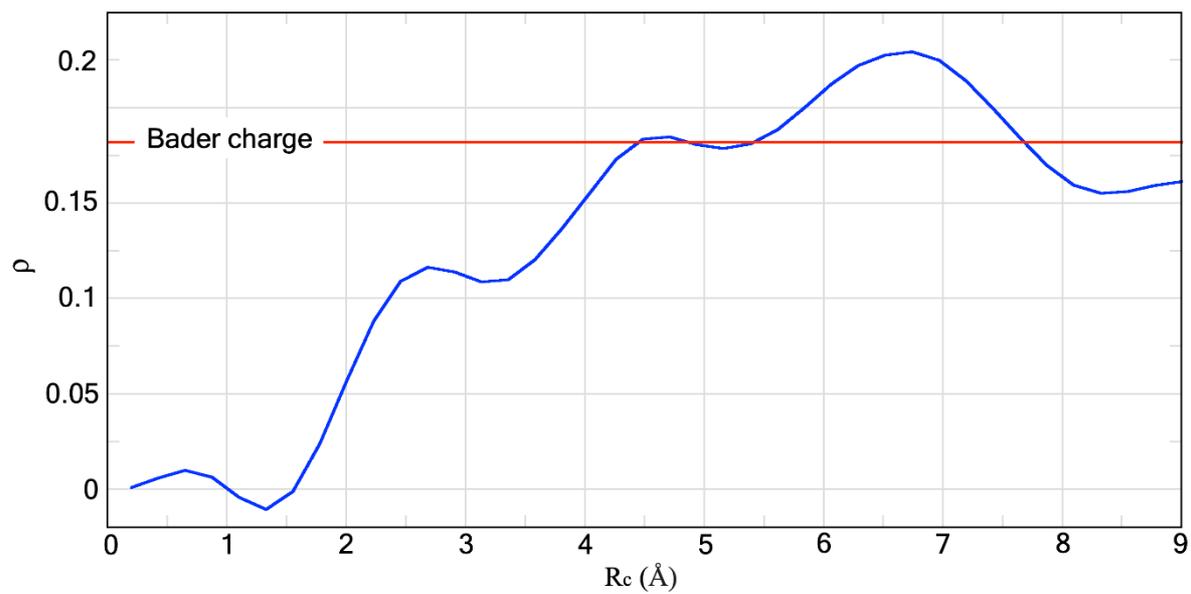

**Fig. S6. Integrated value of transferred charge.**
The value of transferred charge δρ (Fig.S5) integrated outside of sphere Rc with center on the Cs atom. Red line represents the value obtained from Bader charge analysis.

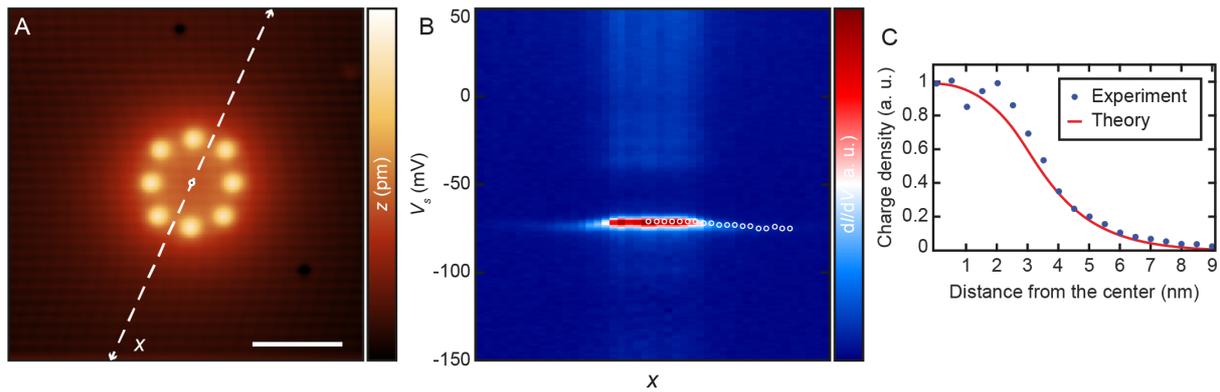

**Fig. S7. Spatially dependence of the *s*-like bound state.**
(**A**) Constant-current STM image of an artificial atom derived from eight Cs atoms arranged in a ring structure (lateral scale: 5 nm, Δ*z* = 300 pm). (**B**) Constant height STS measured sequentially along the dashed white line (*x*) marked in A. Stabilization point is marked with a dot in the middle of the structure. White circles in B indicate the positions of data points presented in C. (**C**) Comparison of the spatially dependent charge density decay for both the experimental data and the calculations for the artificial *s* orbital.

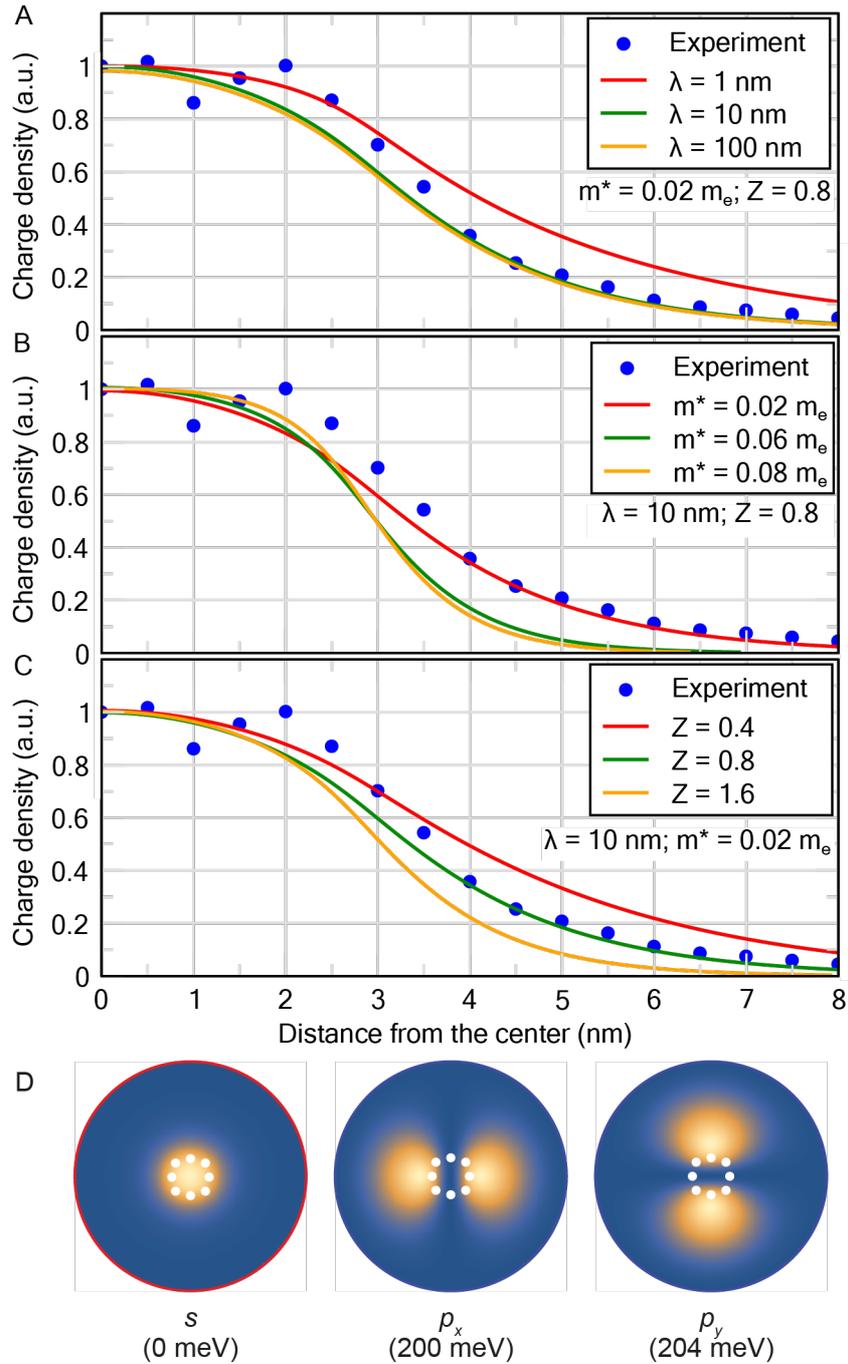

**Fig. S8. Evolution of charge density upon variation of parameters.**
(**A**) Varying screening length λ at fixed parameters $Z = 0.8$ and $m^* = 0.02\,m_e$. (**B**) Varying effective mass $m^*$ at fixed parameters $Z = 0.8$ and $λ = 10$ nm. (**C**) Varying potential charge $Z$ at fixed parameters $m^* = 0.02\,m_e$ and $λ = 10$ nm. (**D**) Three calculated bound states at the given energies and the spatially dependent charge density of the indicated artificial atom. The lowest energy state is identified as an $s$ orbital while higher energy states are identified as $p_x$ and $p_y$ orbitals. The $x$ and $y$ directions correspond to the $[1\bar{1}0]$ and $[001]$ crystallographic directions, respectively.

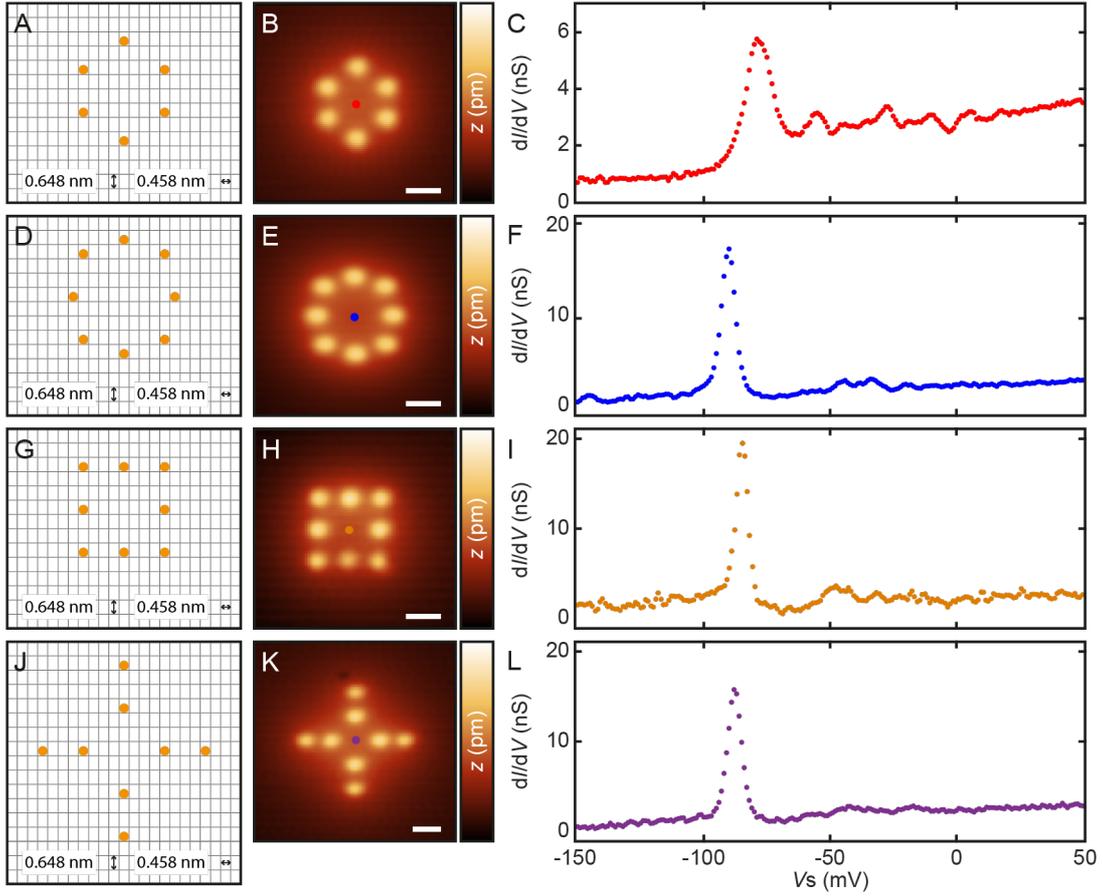

**Fig. S9. Examples of different Cs structures forming an artificial atom.**
(**A**) Model of a hexagonal structure with six Cs atoms. (**B**) Constant-current STM image of an artificial atom derived from six Cs atoms arranged in a hexagonal structure (lateral scale: 2 nm, $\Delta z$ = 300 pm). (**C**) STS measured at the red dot indicated in B, revealing a bound state at $V_S$ = -79 mV. (**D**) Model of a ring structure with eight Cs atoms. (**E**) Constant-current STM image of an artificial atom derived from eight Cs atoms arranged in a ring structure (lateral scale: 2 nm, $\Delta z$ = 300 pm). (**F**) STS measured at the blue dot indicated in E, revealing a bound state at -91 mV. (A, D) Orange dots represent Cs atoms while gray lattice represent Sb sub-lattice of InSb(110) substrate. (**G**) Model of a square structure with eight Cs atoms. (**H**) Constant-current STM image of an artificial atom derived from eight Cs atoms arranged in a square structure (lateral scale: 2 nm, $\Delta z$ = 300 pm). (**I**) STS measured at the orange dot indicated in H, revealing a bound state at $V_S$ = -91 mV. (**J**) Model of a cross structure with eight Cs atoms. (**K**) Constant-current STM image of an artificial atom derived from eight Cs atoms arranged in a cross structure (lateral scale: 2 nm, $\Delta z$ = 300 pm). (**L**) STS measured at the indicated blue dot in K, revealing a bound state at -91 mV. (A, D, G, J) Orange dots represent Cs atoms while gray lattice represent Sb sub-lattice of InSb(110) substrate.

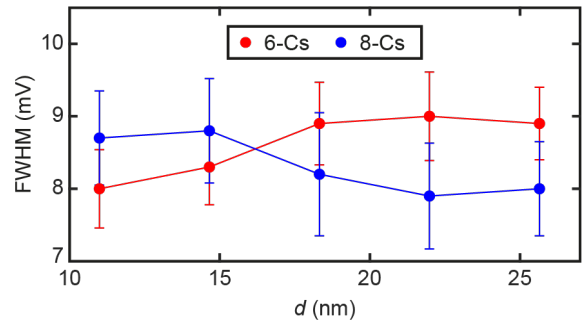

**Fig. S10. Orbital maps of bonding and antibonding states and plot of the FWHM of the bonding state.**
Peak widths of coupled artificial atoms as a function of distance, as determined from Lorentzian fit analyses. Error bars represent uncertainties of the fitting.

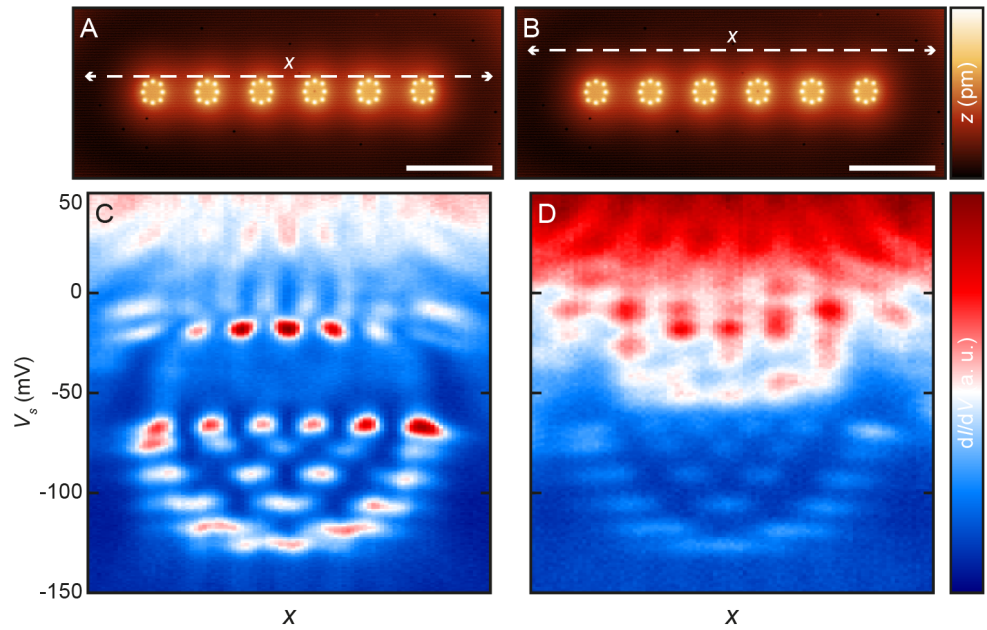

**Fig. S11. Line spectroscopy of the artificial molecular chain.**
(**A-B**) Constant-current STM image of a linear chain composed of six artificial atoms with equal separations $d \approx 12.8$ nm ($V_S = 50$ mV, scale bar: 20 nm, $\Delta z = 300$ pm). (**C-D**) STS spectra measured sequentially along the dashed white line ($x$) marked in A and B, respectively.

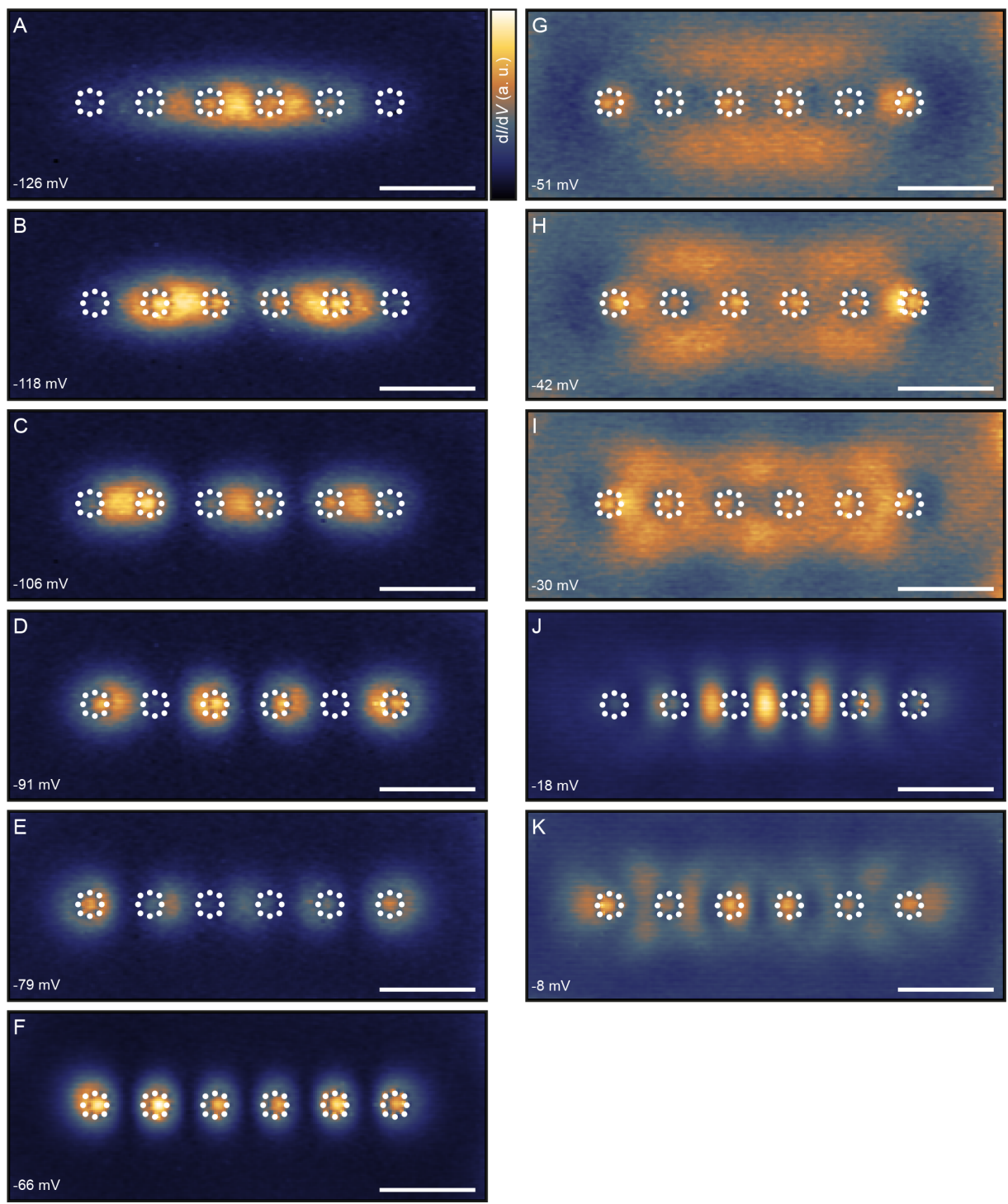

**Fig. S12. Orbital maps of the artificial molecular chain.**
(**A-K**) Orbital maps obtained at the voltages indicated in lower left corner of each image ($z_{offset}$ = -140 pm, $V_{mod}$ = 2 mV, lateral scale: 20 nm). The white circles were added to represent the Cs atoms for clarity. Intensity ranges: (A-F): 2 V; (G-I): 1 V; (J): 3 V; (K): 2 V.

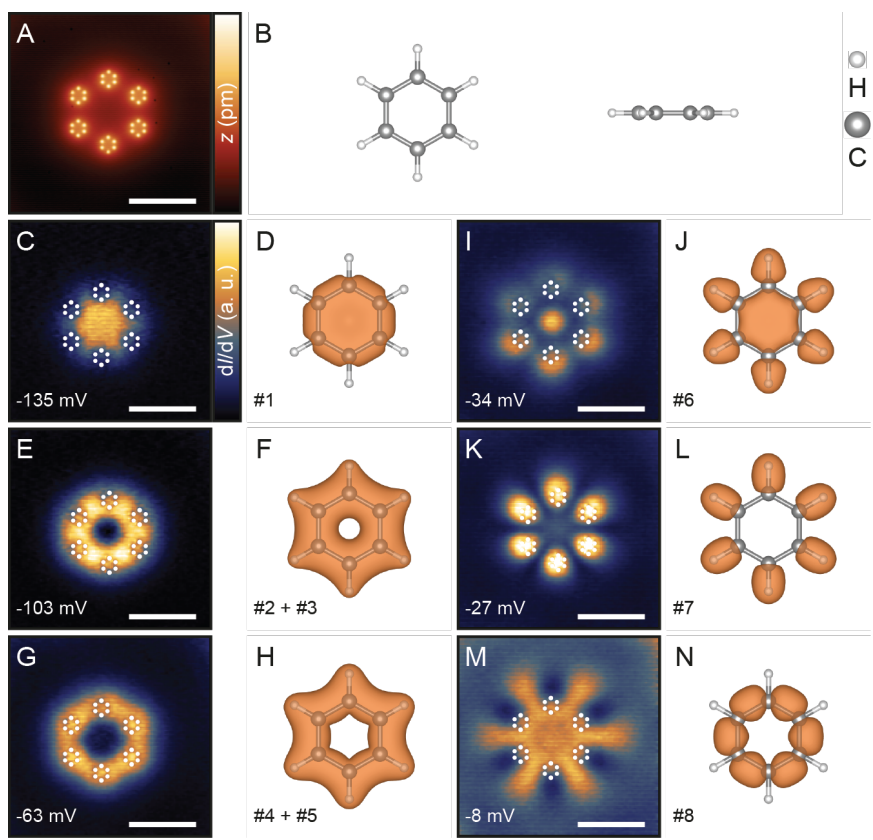

**Fig. S13. Artificial benzene and its orbital structure.**
(**A**) Constant-current STM image of six artificial atoms arranged into a benzene structure with separation of $d \approx 10.5$ nm). (lateral scale: 20 nm, $\Delta z$ = 300 pm). (**B**) The ball-stick model for the benzene molecule in the DFT calculations in the top (left) and side (right) views. (**C, E, G, I, K, M**) Orbital maps obtained at the voltages indicated in the lower left corner of each image ($z_{offset}$ = -140 pm, $V_{mod}$ = 2 mV, lateral scale: 20 nm). The white circles were added to represent the Cs atoms for clarity. (**D, F, H, J, L, N**) The first eight valence molecular orbitals of benzene obtained from DFT calculations. The calculations represent the charge density and include the summed charge densities, where there are degenerate orbitals. The orbital order number is indicated in the lower left corner of each image.

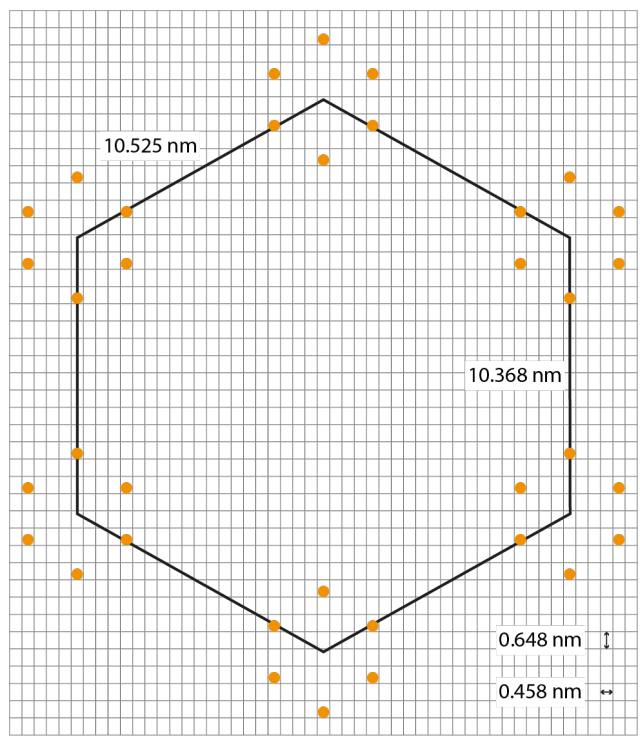

**Fig. S14. Model of the benzene structure.**
Model of a hexagonal structure consisting of six artificial atoms at separations indicated in the figure. Orange dots represent Cs atoms while the gray lattice represents the Sb sub-lattice of the InSb(110) substrate.

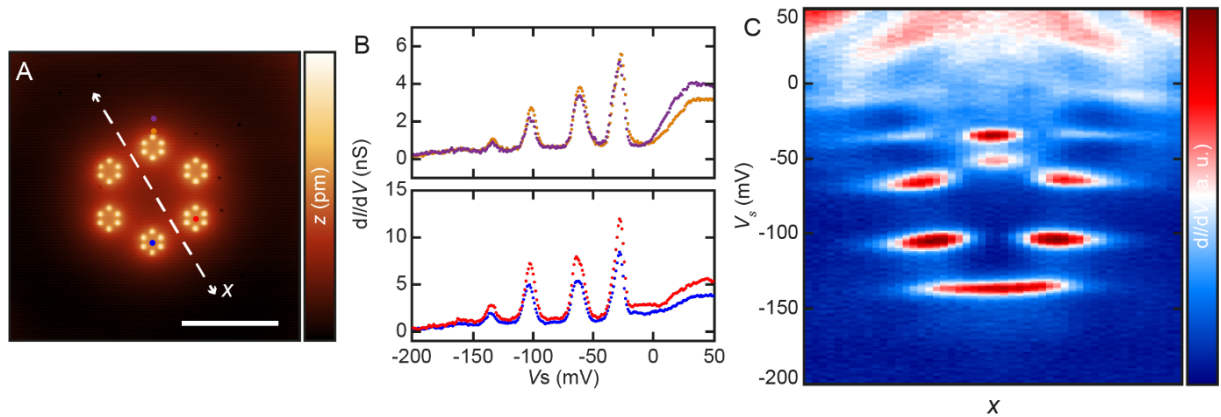

**Fig. S15. STS of the artificial benzene structure.**
(**A**) Constant-current STM image of hexagonal structure composed of six artificial atoms with separations $d \approx 10.5$ nm (lateral scale: 20 nm, $\Delta z = 300$ pm). (**B**) STS measured at the color-coded dots marked in A. (**C**) STS measured sequentially along the dashed white line ($x$) marked in A.

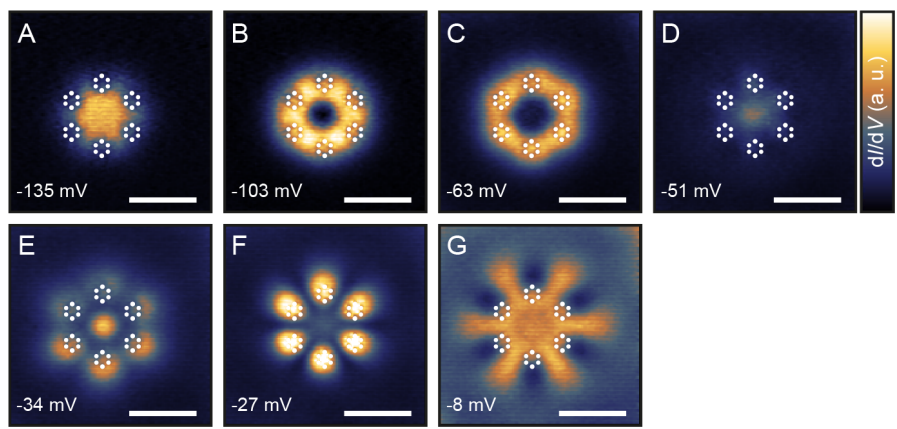

**Fig. S16. Orbital maps of the artificial benzene structure.**
(**A-G**) Orbital maps obtained at the voltages indicated in lower left corner of each image ($z_{offset}$ = -140 pm, $V_{mod}$ = 2 mV, lateral scale: 20 nm). The white circles were added to represent the Cs atoms for clarity. Intensity ranges: (A-F): 2 V; (G): 1 V.

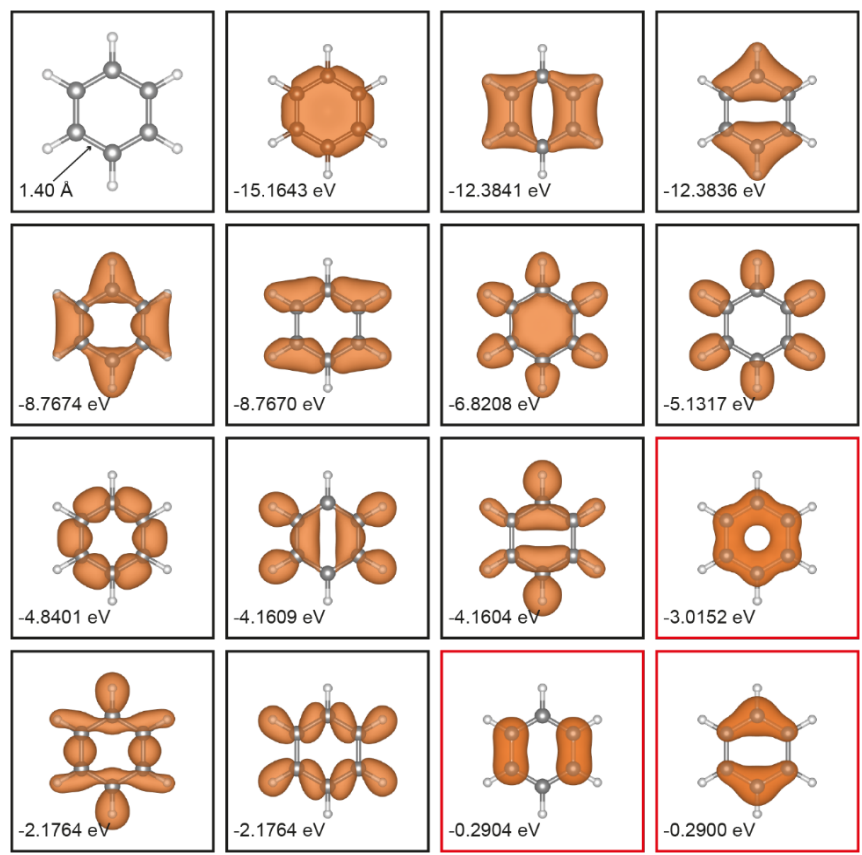

**Fig. S17. Full set of occupied benzene molecular orbitals.**
The VMOs obtained from DFT calculations represent the charge density – the wave function is squared. Energy of each VMO is indicated in lower left corner of each image. VMOs in black boxes are σ orbitals while ones in red boxes are π orbitals. Isosurface value: 0.007.

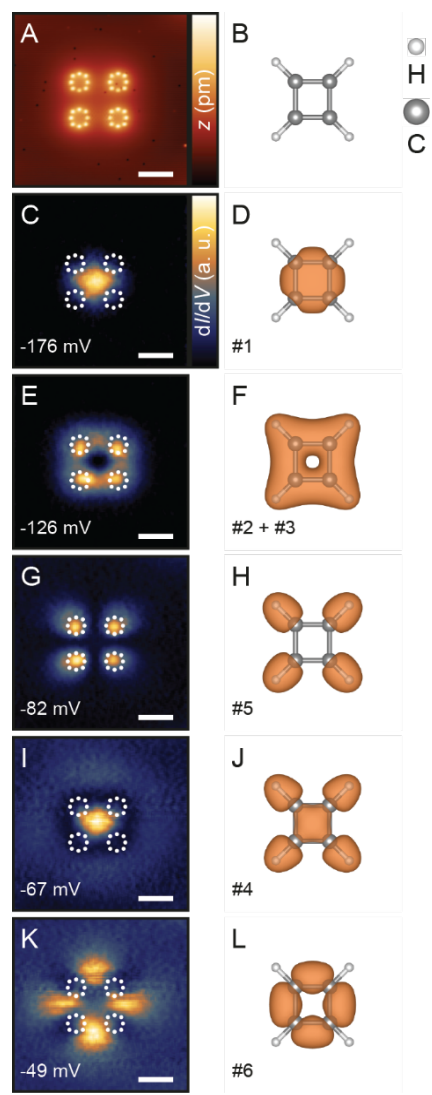

**Fig. S18. Artificial cyclobutadiene and its orbital structure.**
(**A**) Constant-current STM image of four artificial atoms arranged in a cyclobutadiene structure with $d \approx 11$ nm (lateral scale: 10 nm, $\Delta z = 300$ pm). (**B**) The ball-stick model used for the cyclobutadiene molecule in the DFT calculations. (**C, E, G, I, K**) Set of orbital maps obtained at voltages indicated in the lower left corner of each image. The white circles were added to represent the Cs atoms for clarity ($z_{offset} = -100$ pm, $V_{mod} = 1$ mV, lateral scale: 10 nm). (**D, F, H, J, L**) Set of first six cyclobutadiene VMOs obtained from the DFT calculations. The calculations represent the charge density and include the summed charge densities, where there are degenerate orbitals. The orbital order number is indicated in lower left corner of each image.

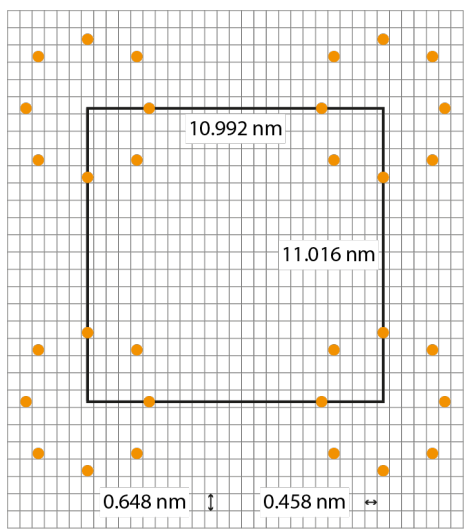

**Fig. S19. Model of the artificial cyclobutadiene structure.**
Model of a square structure consisting of four artificial atoms at separations indicated in the figure. Orange dots represent Cs atoms while the gray lattice represents the Sb sub-lattice of the InSb(110) substrate.

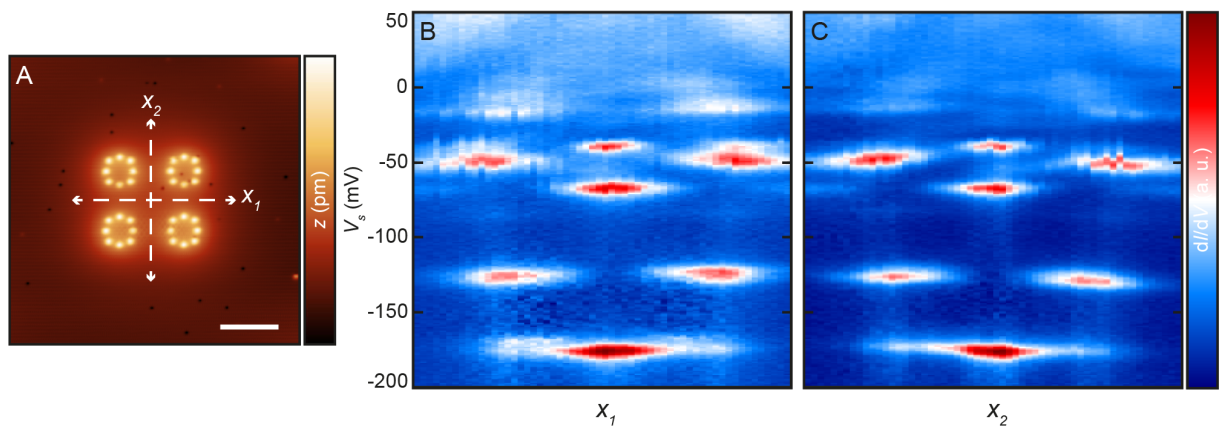

**Fig. S20. Line spectroscopy of the artificial cyclobutadiene structure.**
(**A**) Constant-current STM image of square structure composed of four artificial atoms with separations $d \approx 11$ nm (lateral scale: 10 nm, $\Delta z = 300$ pm). (**B-C**) STS measured sequentially along the dashed white lines ($x_1$ and $x_2$, respectively) marked in A.

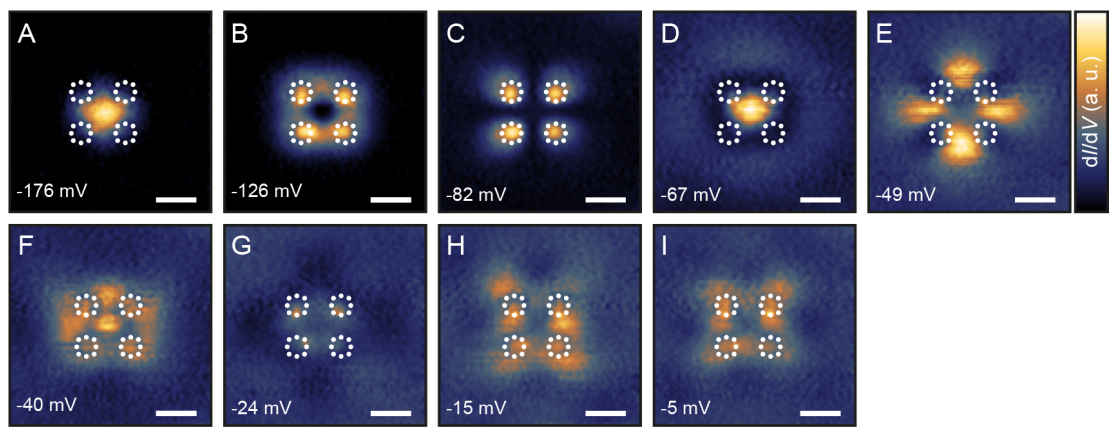

**Fig. S21. Orbital maps of the artificial cyclobutadiene structure.**
(**A-I**) Orbital maps obtained at the voltages indicated in lower left corner of each image ($z_{\text{offset}}$ = -100 pm, $V_{\text{mod}}$ = 1 mV, lateral scale: 10 nm). The white circles were added to represent the Cs atoms for clarity. Intensity ranges: (A): 0.8 V; (B): 1.1 V; (C): 1.3 V; (D-I): 0.8 V.

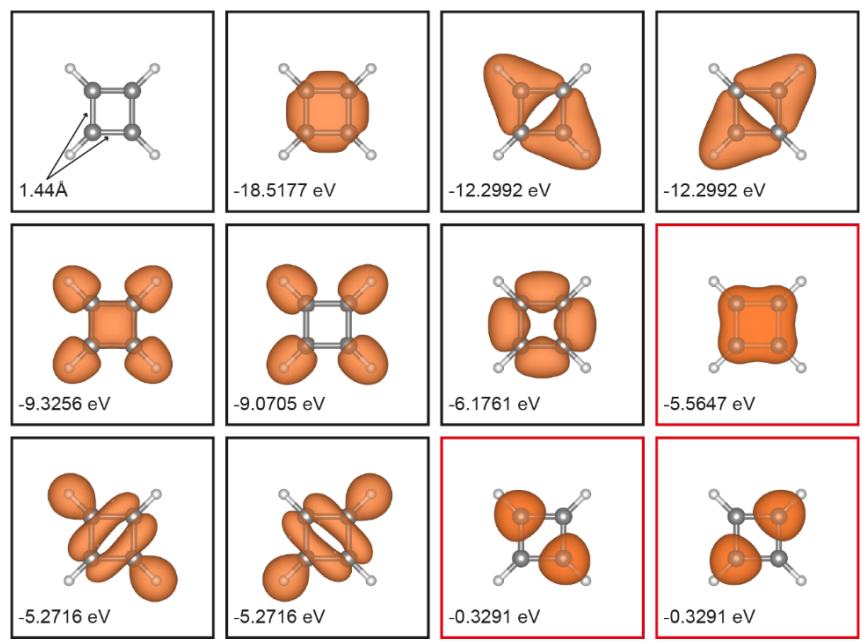

**Fig. S22. Full set of occupied cyclobutadiene molecular orbitals.**
The VMOs obtained from DFT calculations represent the charge density – the wave function is squared. Energy of each VMO is indicated in lower left corner of each image. VMOs in black boxes are σ orbitals while ones in red boxes are π orbitals. Isosurface value: 0.007.

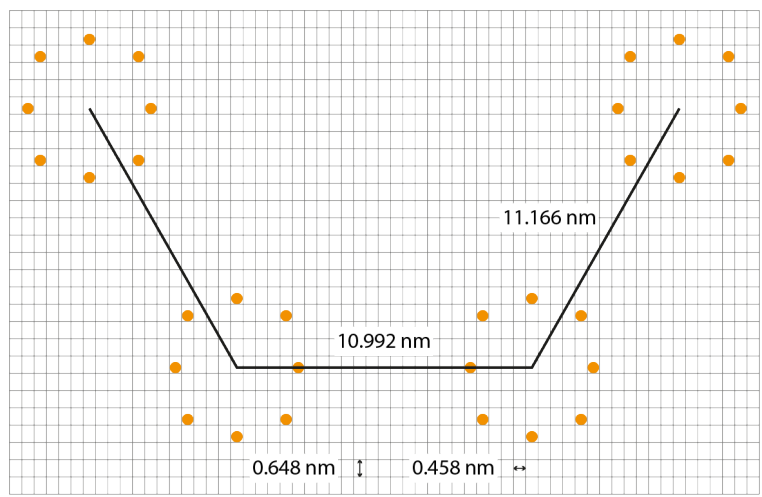

**Fig. S23. Model of the artificial *cis*-butadiene structure.**
Model of a trapezoid structure consisting of four artificial atoms with separations indicated in the figure. Orange dots represent Cs atoms while the gray lattice represents the Sb sub-lattice of the InSb(110) substrate.

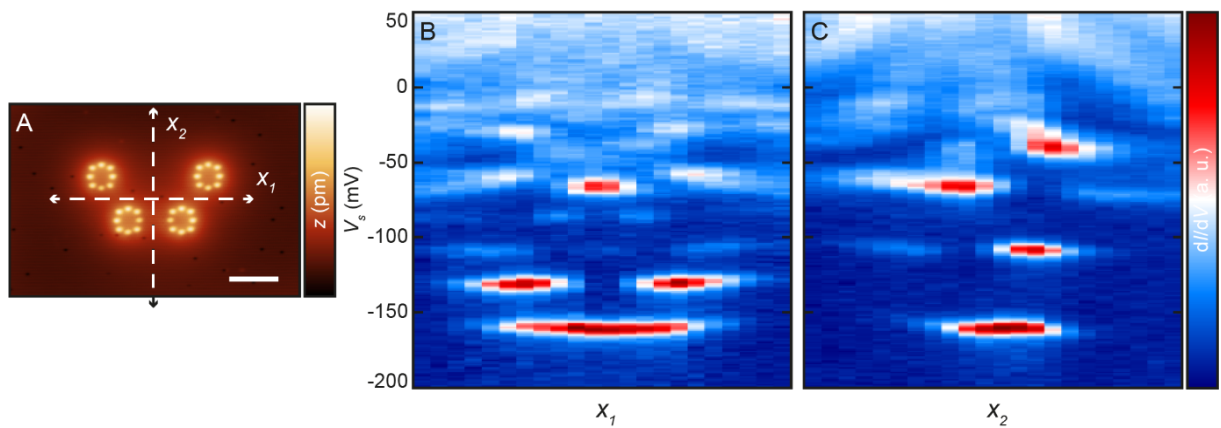

**Fig. S24. Line spectroscopy of the artificial *cis*-butadiene structure.**
(**A**) Constant-current STM image of structure composed of four artificial atoms with separations $d \approx 11$ nm (lateral scale: 10 nm, $\Delta z = 300$ pm). (**B-C**) STS measured sequentially along the dashed white lines ($x_1$ and $x_2$, respectively) marked in A.

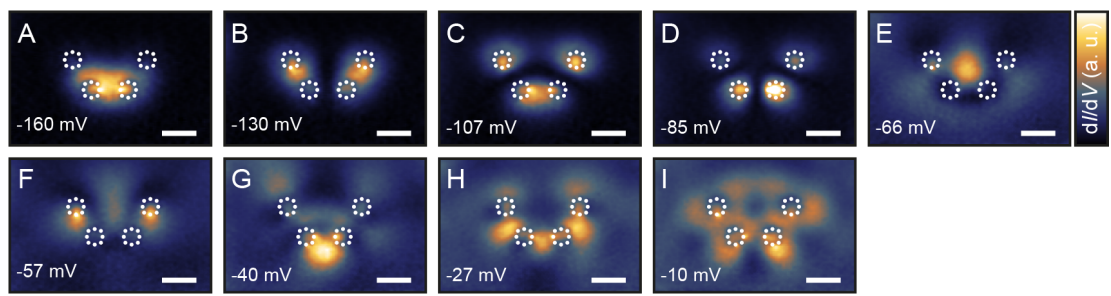

**Fig. S25. Orbital maps of the artificial *cis*-butadiene structure.**
(**A-I**) Orbital maps obtained at the voltages indicated in lower left corner of each image ($z_{offset}$ = -100 pm, $V_{mod}$ = 1 mV, lateral scale: 10 nm). The white circles were added to represent the Cs atoms for clarity. Intensity ranges: (A-D): 1.3 V; (E-I): 0.8 V.

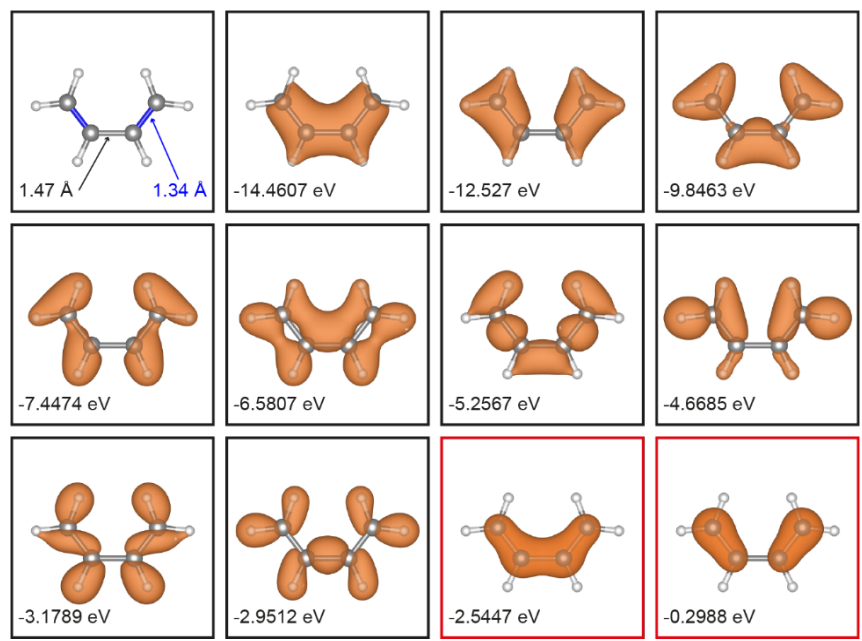

**Fig. S26. Full set of occupied *cis*-butadiene molecular orbitals.**
The VMOs obtained from DFT calculations represent the charge density – the wave function is squared. Energy of each VMO is indicated in lower left corner of each image. VMOs in black boxes are σ orbitals while ones in red boxes are π orbitals. Isosurface value: 0.009.

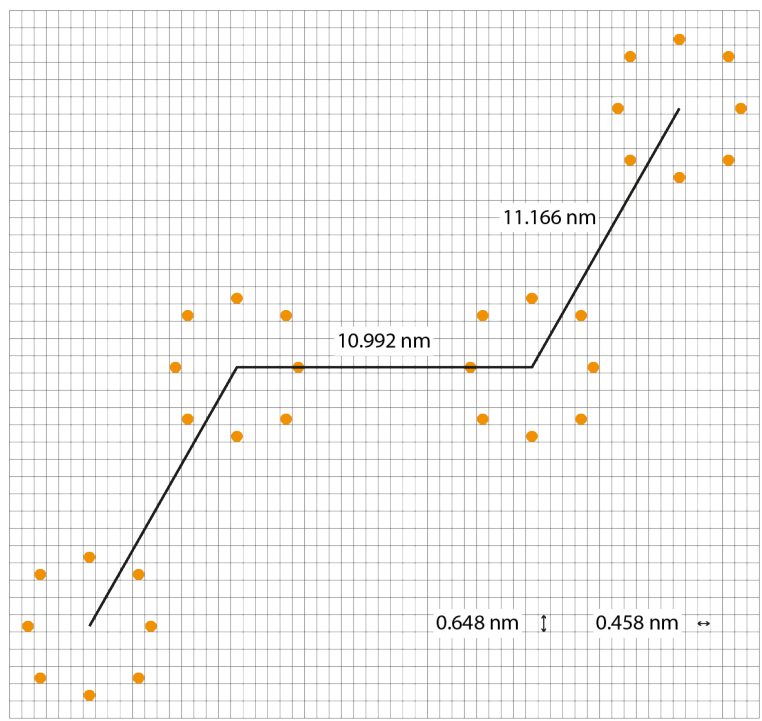

**Fig. S27. Model of the artificial *trans*-butadiene structure.**
Model of a zig-zag structure consisting of four artificial atoms at separations indicated in the figure. Orange dots represent Cs atoms while the gray lattice represents the Sb sub-lattice of the InSb(110) substrate.

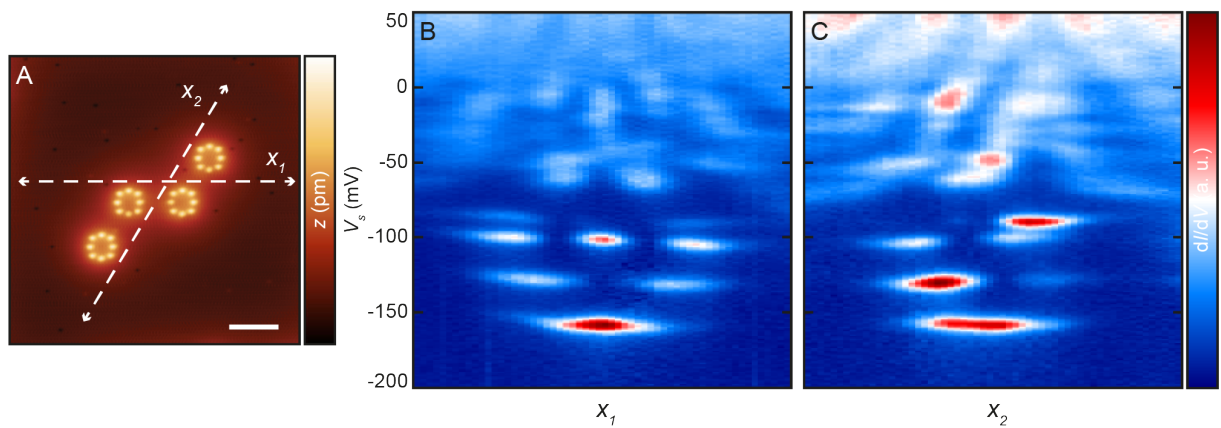

**Fig. S28. Line spectroscopy of the artificial *trans*-butadiene structure.**
(**A**) Constant-current STM image of structure composed of four artificial atoms with separations $d \approx 11$ nm (lateral scale: 10 nm, $\Delta z = 300$ pm). (**B-C**) STS measured sequentially along the dashed white lines ($x_1$ and $x_2$, respectively) marked in A.

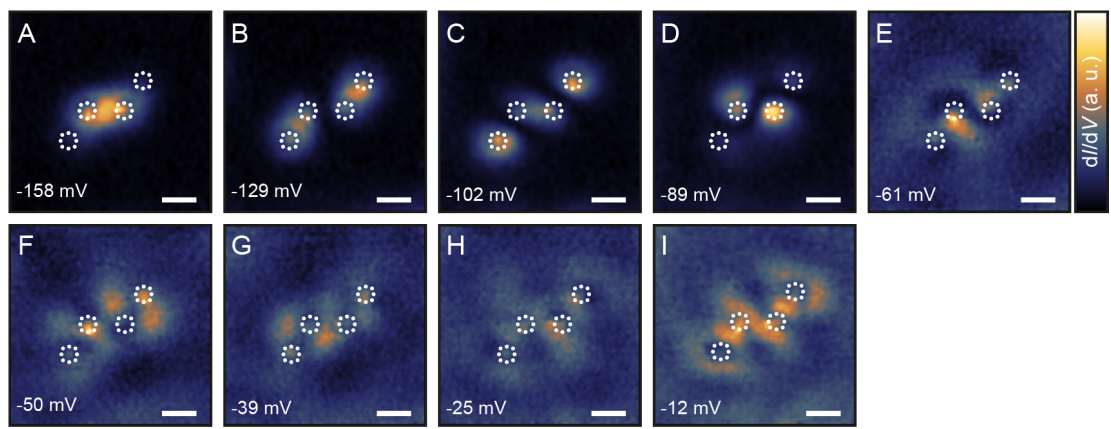

**Fig. S29. Orbital maps of the artificial *trans*-butadiene structure.**
(**A-I**) Orbital maps obtained at the voltages indicated in lower left corner of each image ($z_{\text{offset}}$ = -100 pm, $V_{\text{mod}}$ = 1 mV, lateral scale: 10 nm). The white circles were added to represent the Cs atoms for clarity. Intensity ranges: (A-D): 1.0 V; (E-H): 0.6 V; (I): 0.8 V.

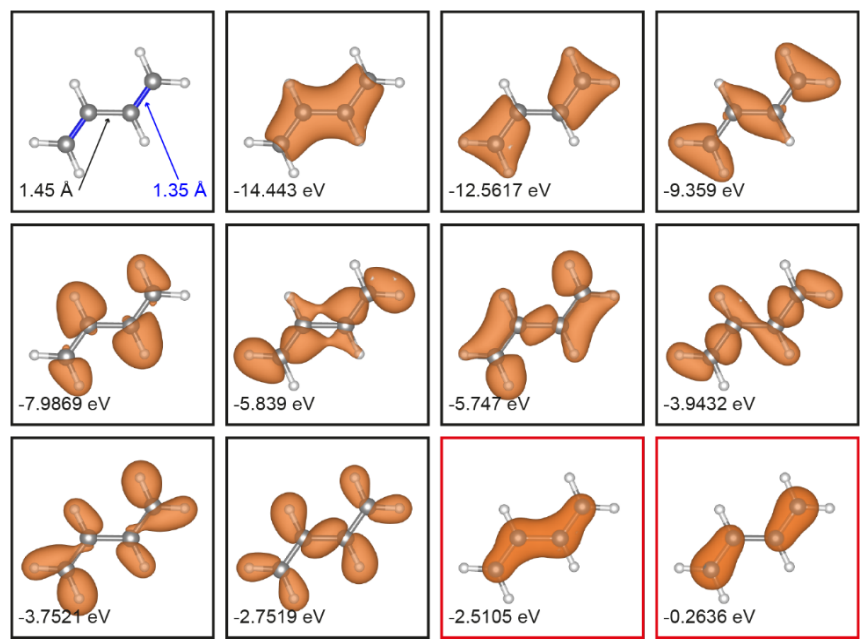

**Fig. S30. Full set of occupied *trans*-butadiene molecular orbitals.**
The VMOs obtained from DFT calculations represent the charge density – the wave function is squared. Energy of each VMO is indicated in lower left corner of each image. VMOs in black boxes are σ orbitals while ones in red boxes are π orbitals. Isosurface value: 0.009.

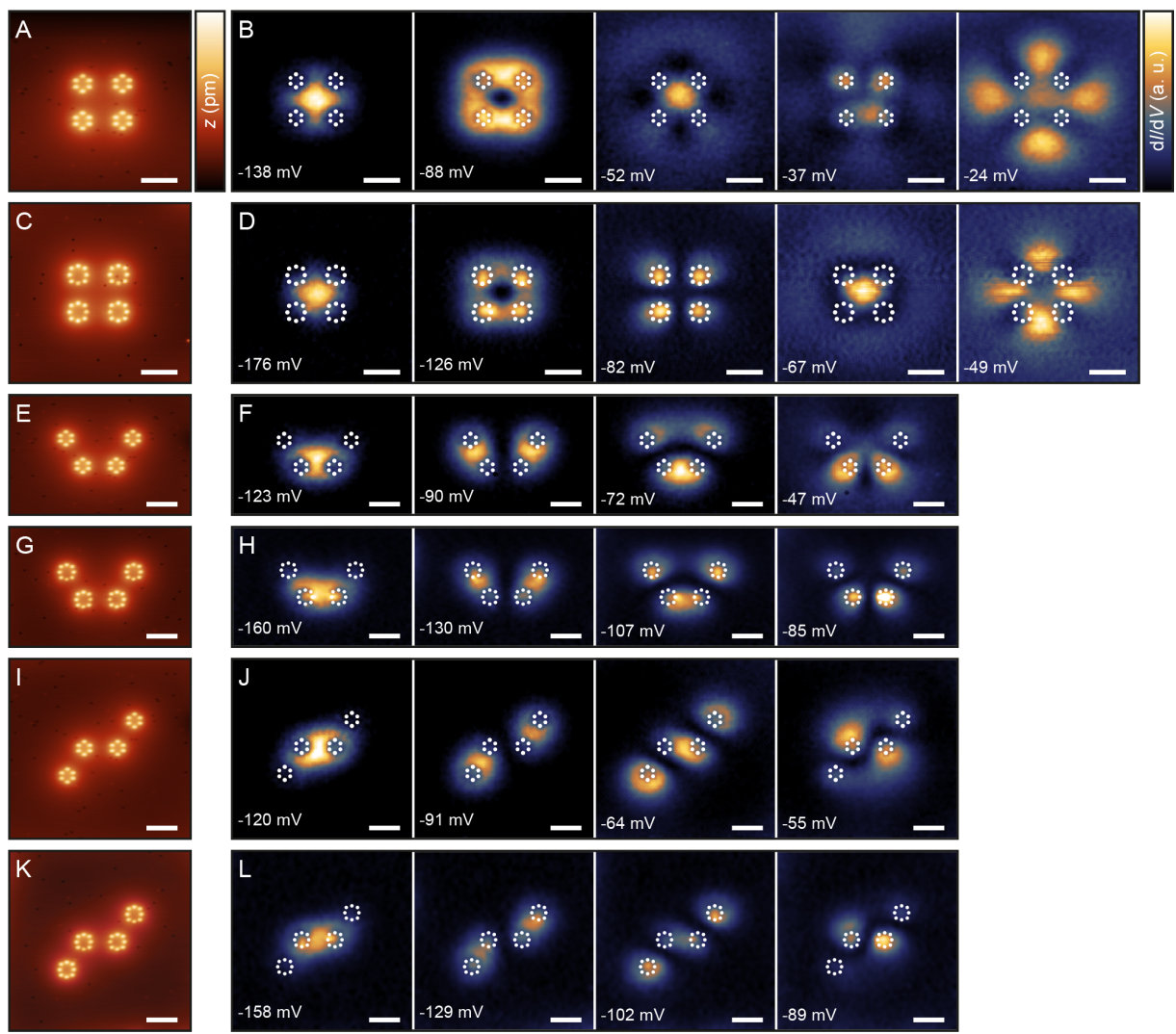

**Fig. S31. Comparison of identical structures built based on 6- and 8-Cs artificial atoms.**
(**A, C, E, G, I, K**) Constant-current STM images of structures composed of four artificial atoms resembling geometries of: (**A, C**) cyclobutadiene; (**E, G**) *cis*-butadiene; (**I, K**) *trans*-butadiene (lateral scale: 10 nm, $\Delta z$ = 300 pm). (**B, D, F, H, J, L**) Orbital maps corresponding to the structures on the left. Maps were obtained at the voltages indicated in lower left corner of each image ($z_{offset}$ (B, F, J) = -140 pm, $z_{offset}$ (D, H, L) = -100 pm $V_{mod}$ = 1 mV, lateral scale: 10 nm). The white circles represent the Cs atoms for clarity. Intensity ranges: (B): (-138 mV to -88 mV): 1.3 V; (-52 mV to -24 mV): 0.6 V; (D): (-176 mV): 0.8 V; (-126 mV): 1.1 V; (-82 mV): 1.3 V; (-67 mV): 0.8 V; (F): (-123 mV to -90 mV): 1.3 V; (-72 mV to -47 mV): 1.1 V; (H): (-160 mV to -85 mV): 1.3 V; (-66 mV to -10 mV): 0.8 V; (J): (-120 mV to -91 mV): 1.3 V; (-64 mV to -55 mV): 1.1 V; (L): (-158 mV to -89 mV): 1.0 V; (-61 mV to -25 mV): 0.6 V; (-12 mV): 0.8 V.

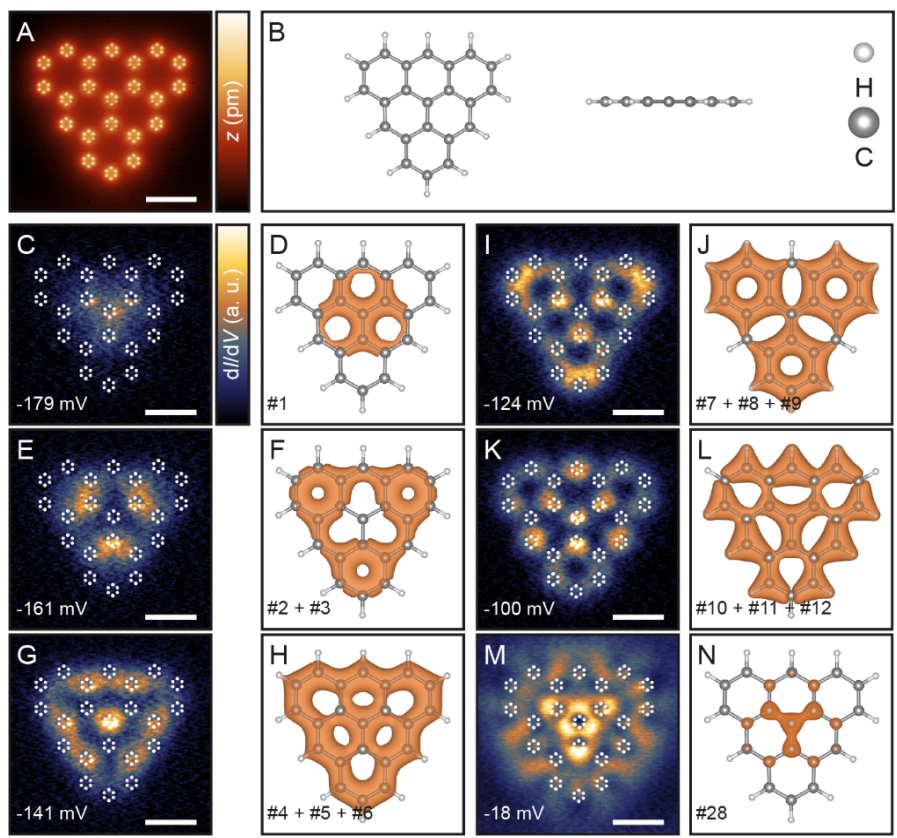

**Fig. S32. Artificial triangulene and comparison to the calculated orbital structure**.
(**A**) Constant-current STM image of twenty-two artificial atoms arranged in a triangulene structure with a separation of $d \approx 10.5$ nm (lateral scale: 20 nm, $\Delta z = 300$ pm). (**B**) The ball-stick model used for the triangulene molecule in the DFT calculations. (**C, E, G, I, K, M**) Orbital maps obtained at the voltages indicated in lower left corner of each image. The white circles were added to represent the Cs atoms for clarity ($z_{offset} = -120$ pm, $V_{mod} = 1$ mV, lateral scale: 20 nm). (**D, F, H, J, L, N**) Set of first twelve triangulene VMOs and the first π VMO (**N**) obtained from DFT calculations. The calculations represent the charge density and include the summed charge densities, where there are VMOs which are degenerate or close to each other. VMO order number is indicated in lower left corner of each image.

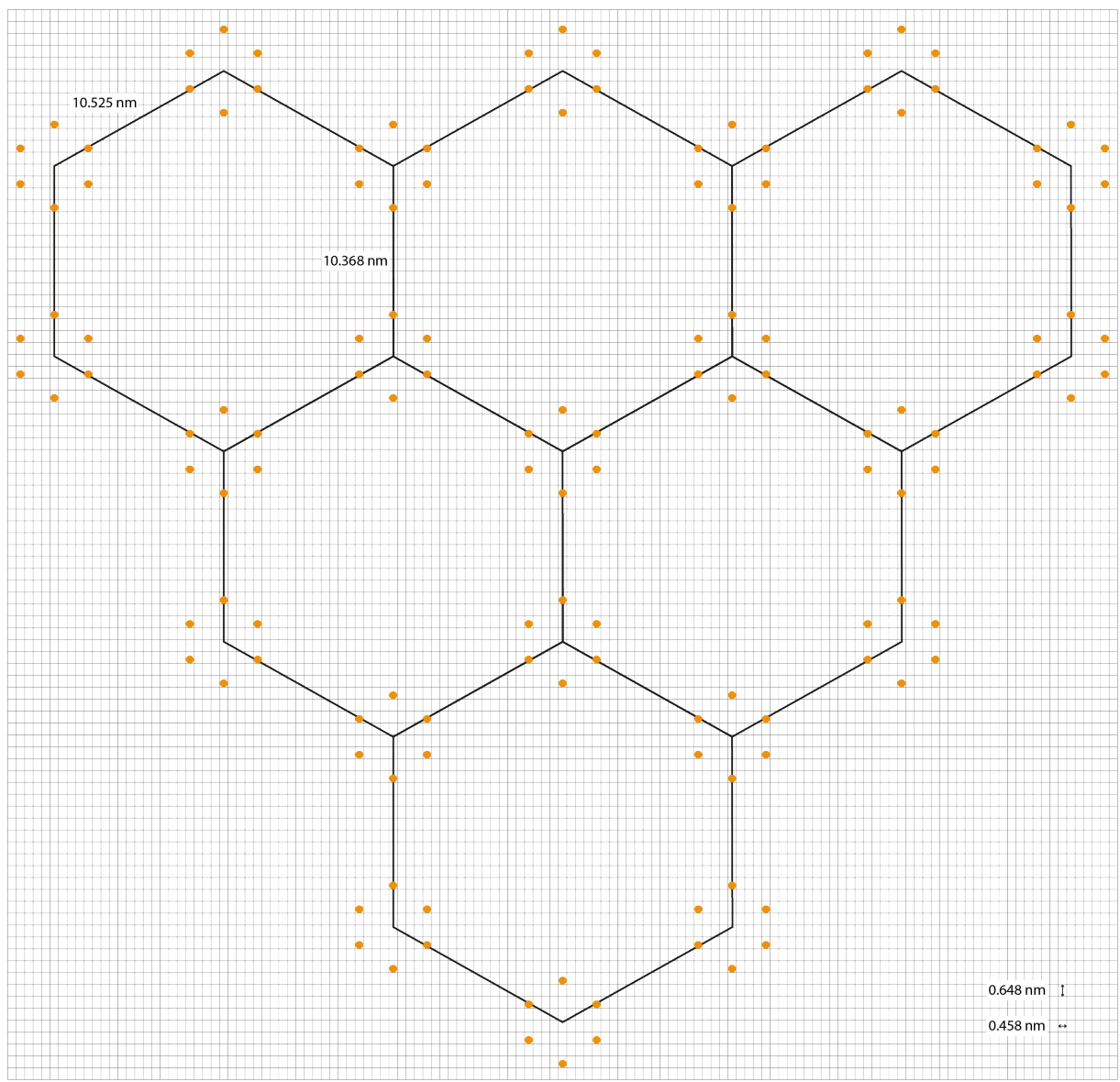

**Fig. S33. Model of the artificial triangulene structure.**
Model of a triangular structure consisting of twenty-two artificial atoms at separations indicated in the figure. The orange dots represent Cs atoms while the gray lattice represents the Sb sub-lattice of the InSb(110) substrate.

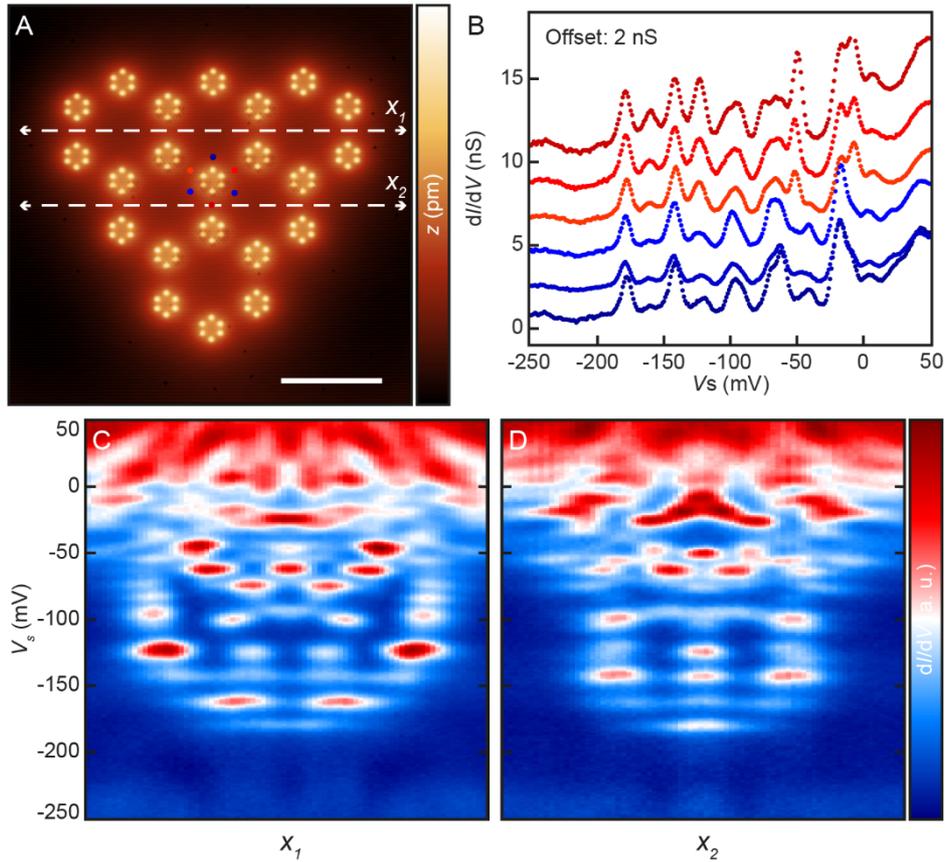

**Fig. S34. STS of the artificial triangulene structure.**
(**A**) Constant-current STM image of 22 artificial atoms arranged in a triangulene structure with a separation of $d \approx 10.5$ nm (lateral scale: 20 nm, $\Delta z = 300$ pm). (**B**) STS measured at the positions marked as colored dots in (A). The spectra are offset by 2nS for clarity. (**C-D**) STS measured sequentially along the dashed white lines ($x_1$ and $x_2$, respectively) marked in (A).

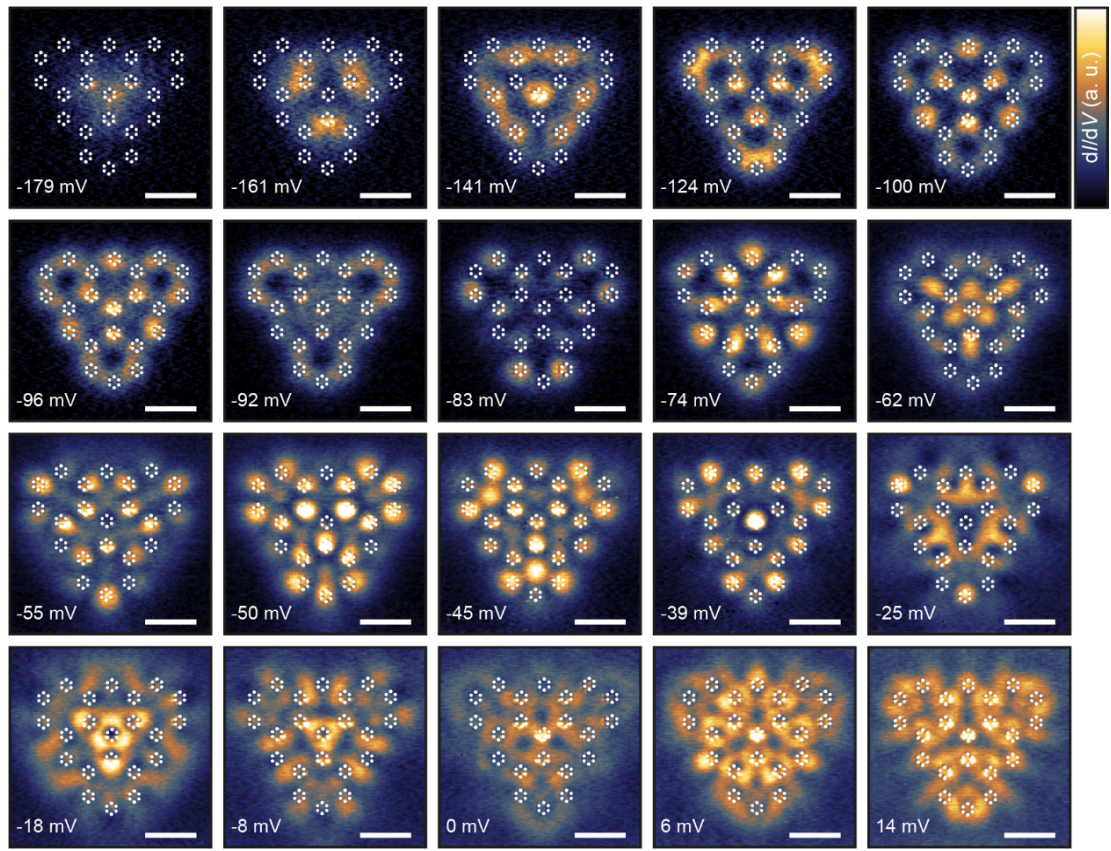

**Fig. S35. Orbital maps of the artificial triangulene structure.**
Orbital maps obtained at the voltages indicated in lower left corner of each image ($z_{\text{offset}}$ = -120 pm, $V_{\text{mod}}$ = 1 mV, lateral scale: 20 nm). The white circles were added to represent the Cs atoms for clarity. Intensity range: 0.5 V.

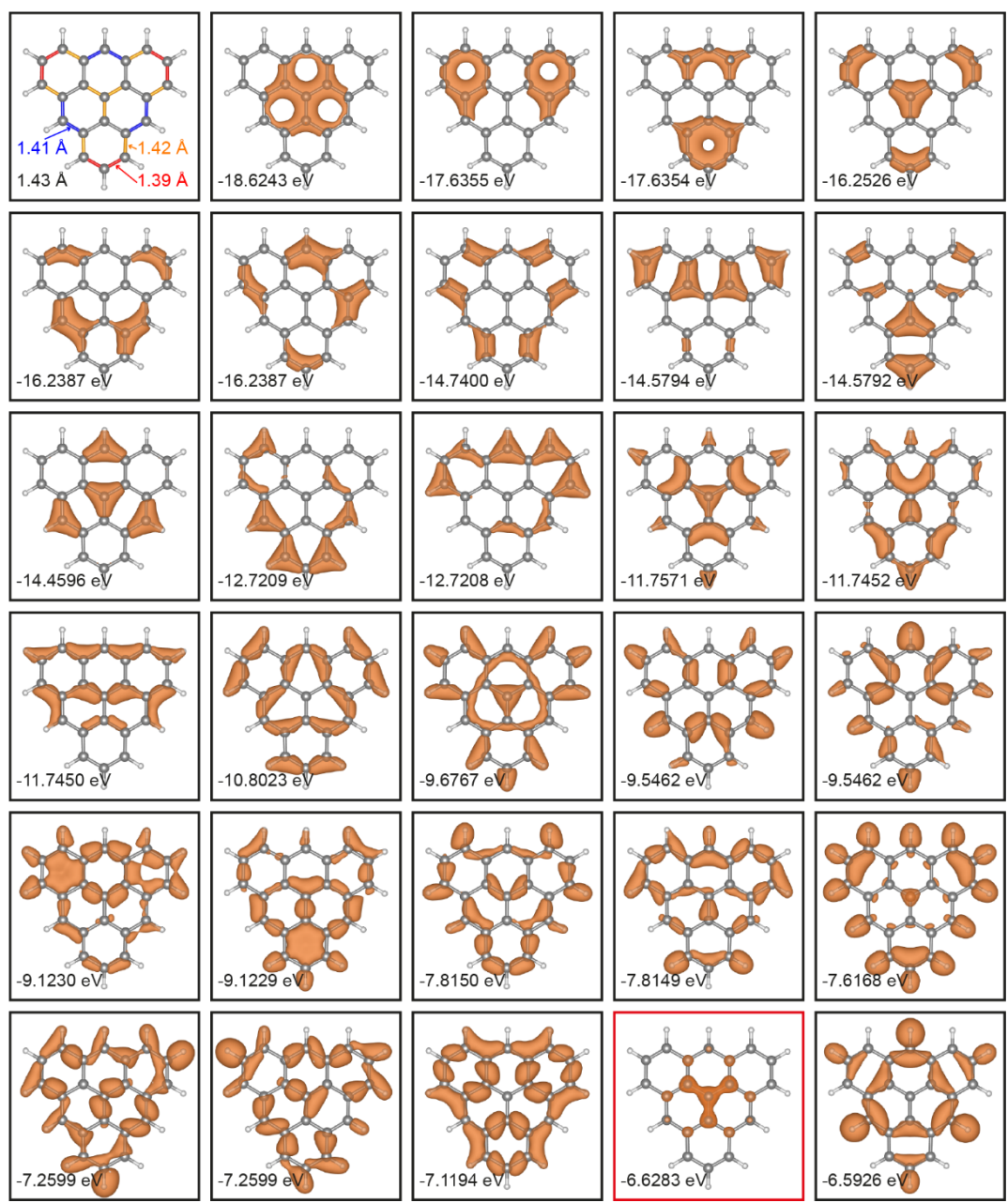

**Fig. S36. The first 29 occupied molecular orbitals of triangulene.**
The VMOs obtained from DFT calculations represent the charge density – the wave function is squared. Energy of each VMO is indicated in lower left corner of each image. VMOs in black boxes are σ orbitals while ones in red boxes are π orbitals. Isosurface values: (-21.9551 eV to -15.0758 eV): 0.007; (-14.1331 eV to -11.1457 eV): 0.005; (-10.9476 eV to -10.4502 eV): 0.003; (-9.9591 eV): 0.007; (-9.9234 eV and -9.6263 eV): 0.003.

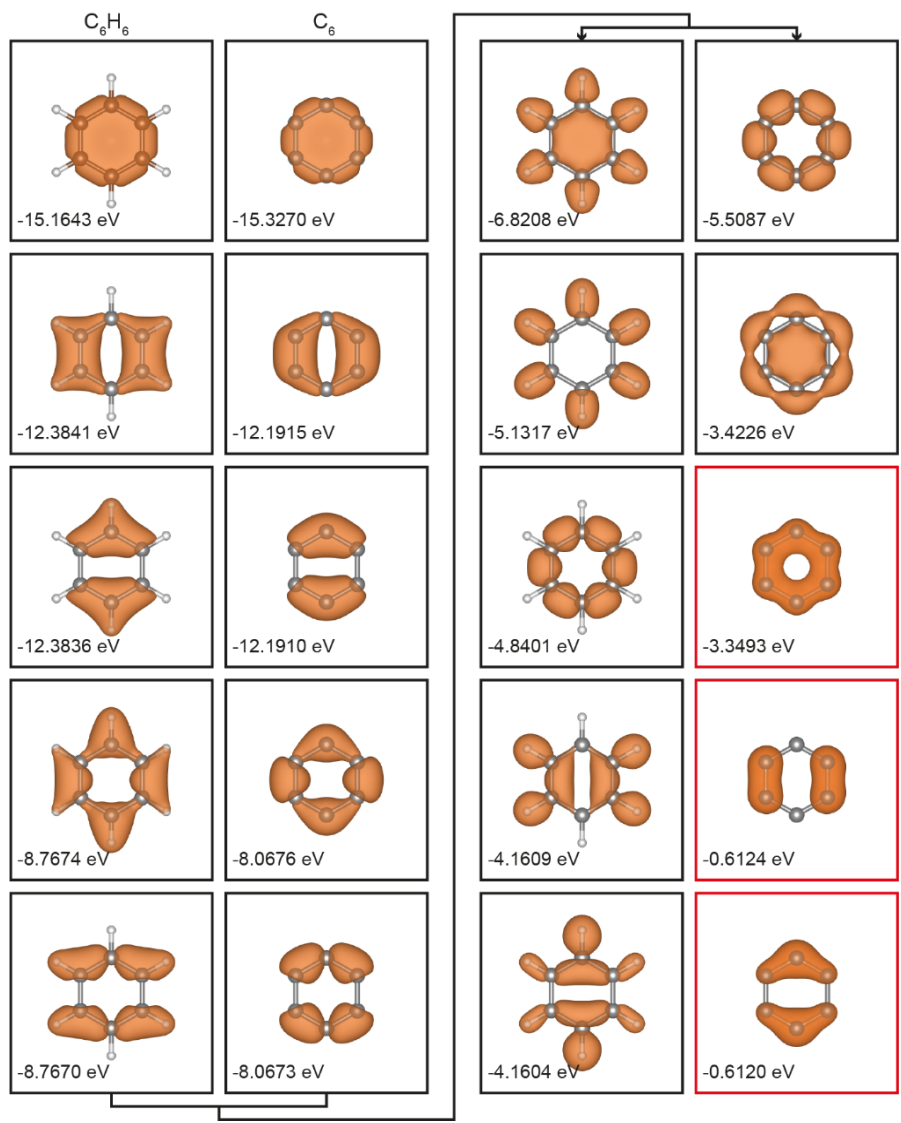

**Fig. S37. Comparison between molecular orbitals of $C_6H_6$ and $C_6$.**
The VMOs obtained from DFT calculations represent the charge density – the wave function is squared. Energy of each VMO is indicated in lower left corner of each image. VMOs in black boxes are σ orbitals while ones in red boxes are π orbitals. Isosurface value: 0.007.

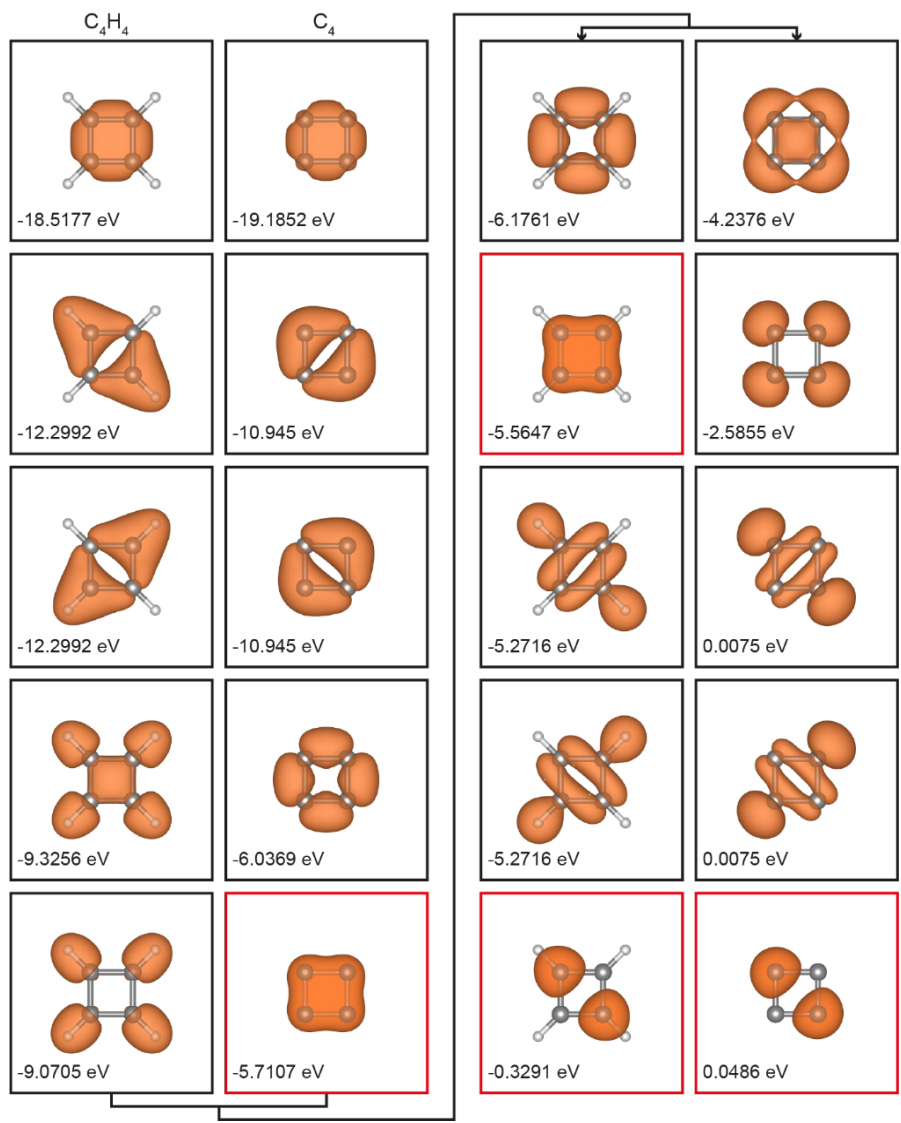

**Fig. S38. Comparison between molecular orbitals of $C_4H_4$ and $C_4$.**
The VMOs obtained from DFT calculations represent the charge density – the wave function is squared. Energy of each VMO is indicated in lower left corner of each image. VMOs in black boxes are σ orbitals while ones in red boxes are π orbitals. Isosurface value: 0.007.

**Table S1: Relative energies of Cs ring structure for different parameters in Eq. S2.**

|  | $\Delta E_1$ (meV) | $\Delta E_2$ (meV) | $\Delta E_3$ (meV) |
|---|---|---|---|
| $\lambda = 1$ nm<br>$Z = 0.8$, $m^* = 0.02$ $m_e$ | 107 | 108 | 208 |
| $\lambda = 10$ nm<br>$Z = 0.8$, $m^* = 0.02$ $m_e$ | 200 | 204 | 300 |
| $\lambda = 100$ nm<br>$Z = 0.8$, $m^* = 0.02$ $m_e$ | 210 | 214 | 314 |
| $m^* = 0.02$ $m_e$<br>$Z = 0.8$, $\lambda = 10$ nm | 200 | 204 | 300 |
| $m^* = 0.06$ $m_e$<br>$Z = 0.8$, $\lambda = 10$ nm | 147 | 156 | 265 |
| $m^* = 0.08$ $m_e$<br>$Z = 0.8$, $\lambda = 10$ nm | 124 | 133 | 266 |
| $Z = 0.4$<br>$m^* = 0.02$ $m_e$, $\lambda = 10$ nm | 127 | 128 | 228 |
| $Z = 0.8$<br>$m^* = 0.02$ $m_e$, $\lambda = 10$ nm | 200 | 204 | 300 |
| $Z = 1.6$<br>$m^* = 0.02$ $m_e$, $\lambda = 10$ nm | 347 | 362 | 528 |